\documentclass[useAMS,usenatbib,onecolumn]{mn2e}
\usepackage{graphicx}
\usepackage{bm}
\usepackage{epsfig}
\usepackage{amssymb, ulem}
\usepackage{amsmath}
\usepackage{empheq}
\usepackage{amsfonts}
\usepackage{cleveref}
\usepackage{courier}
\usepackage{bm}
\usepackage{booktabs}
\usepackage{tabularx}
\usepackage[makeroom]{cancel}

\usepackage[utf8]{inputenc}

\newcommand{\bea}{\begin{eqnarray}}
\newcommand{\eea}{\end{eqnarray}}

\newcommand{\beq}{\begin{equation}}
\newcommand{\eeq}{\end{equation}}

\newcommand{\simless}[0]{\mathbin{\lower 3pt\hbox
   {$\rlap{\raise 5pt\hbox{$\char'074$}}\mathchar"7218$}}}
\newcommand{\simgreat}[0]{\mathbin{\lower 3pt\hbox
   {$\rlap{\raise 5pt\hbox{$\char'076$}}\mathchar"7218$}}}

\newcommand{\figref}[1]{figure \ref{#1}}
\newcommand{\figrefs}[1]{figures \ref{#1}}
\newcommand{\figrefbare}[1]{\ref{#1}}
\newcommand{\capfigref}[1]{Figure \ref{#1}}
\newcommand{\capfigrefs}[1]{Figures \ref{#1}}
\newcommand{\eqnref}[1]{eq. (\ref{#1})} 
\newcommand{\eqnrefs}[1]{eqs. (\ref{#1})} 
\newcommand{\eqnrefbare}[1]{(\ref{#1})} 
\newcommand{\capeqnref}[1]{Equation (\ref{#1})}
\newcommand{\capeqnrefs}[1]{Equations (\ref{#1})}

\graphicspath{{}{figures/}}

\title[Non-linear density-velocity dynamics in $f(R)$]{Non-linear density-velocity dynamics in $f(R)$ gravity from spherical collapse. }
\author[Sharvari Nadkarni-Ghosh and Sandip Chowdhury]{Sharvari Nadkarni-Ghosh $^{1}$\thanks{E-mail:nsharvari@gmail.com, sharvari@iitk.ac.in (Corresponding Author)} and Sandip Chowdhury$^{1}$\thanks{E-mail:sandipc@iitk.ac.in} \\
$^{1}$Department of Physics, I.I.T. Kanpur, Kanpur, U.P. 208016 India  }

\begin{document}
\date{}

\maketitle
\begin{abstract}
We investigate the joint density-velocity evolution in $f(R)$ gravity using smooth, compensated spherical top-hats as a proxy for the non-linear regime. Using the Hu-Sawicki model as a working example, we solve the coupled continuity, Euler and Einstein equations  using an iterative hybrid Lagrangian-Eulerian scheme. The novel aspect of this scheme is that the metric potentials are solved for analytically in the Eulerian frame. The evolution is assumed to follow GR at very early epochs and switches to $f(R)$ at a pre-determined epoch. Choosing the `switching epoch' too early is computationally expensive because of high frequency oscillations; choosing it too late potentially destroys consistency with $\Lambda$CDM. To make an informed choice, we perform an eigenvalue analysis of the background model which gives a ballpark estimate of the magnitude of oscillations. There are two length scales in the problem: the comoving Compton wavelength of the associated scalar field and the width of the top-hat.  The evolution is determined by their ratio. When the ratio is large, the evolution is scale-independent and the density-velocity divergence relation (DVDR) is unique. When the ratio is small, the evolution is very close to GR, except for the formation of a spike near the top-hat edge, a feature which has been noted in earlier literature. We are able to qualitatively explain this feature in terms of the analytic solution for the metric potentials, in the absence of the chameleon mechanism. In the intermediate regime, the evolution is profile-dependent and no unique DVDR exists.
\end{abstract}

\section{Introduction}
\label{sec:intro}
Observations of type IA supernovae have established that our Universe is undergoing a phase of accelerated expansion. The best explanation so far, is that this acceleration is due to an additional constant energy density component - called the cosmological constant ($\Lambda$). On galactic scales, observations of galaxy rotation curves have established the need for a non-baryonic matter component. In the standard $\Lambda$CDM model, our Universe today is made up of about 68\% of $\Lambda$, 28\% of dark matter and 4\% baryonic or visible matter \citep{planck_collaboration_planck_2020}. While this model agrees remarkably well with a host of other observational constraints, there are still some glaring issues. The physical origin of $\Lambda$ and dark matter are still unknown. On the observational side, recent measurements indicate that CMB estimates of the Hubble constant ($H_0$) and the amplitude of fluctuations ($\sigma_8$) 
are in disagreement with other probes of the same quantities (for e.g., \citealt{Macaulay_2013,Di_Valentino_2021}). Thus, the need arises to explore models beyond $\Lambda$CDM. There are two broad avenues that have been considered. One path continues to assume that the observed acceleration is due to an additional `dark energy' component and a plethora of phenomenological models have been explored. The other path assumes that there is no additional energy density, but instead postulates that Einstein's General Relativity (GR) is incorrect and needs modification (see reviews by \citealt{sotiriou_fr_2010,defelice_fR_2010,clifton_modified_2012,joyce_beyond_2015, joyce_dark_2016,Nojiri_2017} for various aspects of this subject). Current and future cosmological surveys aim to constrain modified gravity (MG hereafter) models through a host of observables. The number counts of clusters, their density and velocity profiles, the splashback radius and weak lensing maps are some of the important cluster-scale probes that will help constrain model parameters. Similarly, redshift-space distortions that give estimates of galaxy peculiar velocities also provide complementary constraints (see the recent review by \citealt{baker_novel_2021}).  

In standard GR, the perturbations on sub-horizon scales, in the single stream limit, obey the coupled continuity, Euler and Poisson equations for the fractional overdensity $\delta$, the peculiar velocity ${\bf v}$ and the peculiar gravitational potential $\Psi$. In standard GR, in the linear regime, the density and peculiar velocity are related through the continuity equation as 
\beq 
\Theta = - f(\Omega_m) \delta
\label{dvdrlinear}
\eeq
where $\Theta = H^{-1} \nabla_r \cdot {\bf v}$ is the scaled divergence of the peculiar velocity w.r.t. the physical coordinate ${\bf r}$. $f = d \ln \delta/ d \ln a$ is defined to be the linear growth rate and is usually expressed as $f(\Omega_m) = \Omega_m^\gamma$, where $\gamma$ is called the growth index and is sensitive to the cosmological model. Indeed, one of the main aims of surveys like DESI\footnote{https://www.desi.lbl.gov/} (e.g.,\citealt{alam_testing_2020}), 
SDSS\footnote{https://www.sdss.org} (e.g., \citealt{alam_SDSS_2017}) and Euclid\footnote{http://sci.esa.int/euclid/.} (e.g., \citealt{amendola_2018}) is to test the above relation and constrain $f$. This relation has also been used extensively to constrain the growth rate from local flow measurements (\citealt{nusser_new_2012,nusser_velocity-density_2017,lilow_constrained_2021}) or for measuring the velocity of the local group (e.g., \citealt{nusser_recovery_2014}). However, this relation breaks down on non-linear scales and this deviation can contribute to the error budget in the determination of $f$ \citep{nusser_new_2012}. Velocities provide a bias-independent measure of the underlying density field and hence the non-linear velocity-gravity relation has been of interest since the early 1990s.  Analytic estimates were given based on perturbation theory both in Eulerian and Lagrangian frames (for e.g., \citealt{bertschinger_recovering_1989,nusser_cosmological_1991,bernardeau_quasi-gaussian_1992,gramann_second-order_1993,gramann_improved_1993,chodorowski_large-scale_1997,chodorowski_weakly_1997, susperregi_cosmic_1997,chodorowski_recovery_1998,kitaura_estimating_2012}) and were validated and extended by numerical simulations (for e.g, \citealt{mancinelli_local_1995,mancinelli_nonlinear_1993,bernardeau_new_1996,bernardeau_non-linearity_1999,zaroubi_wiener_1999,kudlicki_reconstructing_2000,ciecielg_gaussianity_2003,scoccimarro_redshift-space_2004,colombi_cosmic_2007, hahn_properties_2015}). Fitting forms for $\Lambda$CDM were provided based on spherical collapse (\citealt{bilicki_velocity-density_2008}) and extended to other dark energy models  and triaxial collapse (\citealt{nadkarni-ghosh_non-linear_2013,nadkarni-ghosh_phase_2016,mandal_one-point_2020}). 

All of the above mentioned investigations of the density-velocity divergence relation (DVDR),  also sometimes called the velocity-gravity relation have assumed standard GR. One of the important features of standard GR is that the `Newtonian' gravitational potential $\Psi$, which dictates particle dynamics and appears in the Euler equation is the same as the potential $\Phi$ that governs the spatial curvature and is related to the density through the Poisson equation. This equality is violated in most MG models and an extra equation is necessary to evolve the two potentials simultaneously. This changes the fundamental structure of the evolution equations and the interplay between the continuity and Euler equations which couples the density and velocity is different for MG as compared to GR (see figures 1 of \citealt{uzan_acceleration_2007} and \citealt{yong_step_2009} for nice schematics). 

In this paper, we aim to understand the joint density-velocity dynamics using the spherical collapse model as a proxy for the non-linear regime. In the linear regime, the DVDR arises from imposing the `slaving condition'. It implies that the velocity field is proportional to the acceleration field and is such that there are no perturbations at the big bang epoch. In first order Eulerian perturbation theory, this is equivalent to ignoring the decaying mode in the linear solution for $\delta$ \citep{Peebles80} and in Lagrangian perturbation theory, it is embedded in the Zeldovich approximation \citep{zeldovich_gravitational_1970,buchert_92}. In \cite{nadkarni-ghosh_non-linear_2013}, it was shown that it is possible to extend this same idea to the non-linear regime to obtain the non-linear DVDR. In a spherically symmetric system, the condition of no perturbations at the big bang, sets a specific relation between the non-linear $\delta$ and $\Theta$ at any epoch. 
This relation traces out a special curve in the two dimensional $\delta-\Theta$ phase space, which was termed as the `Zeldovich curve' and it was shown that this curve is an invariant set of the dynamical system given by the joint continuity and Euler equations, restricted to spherical symmetry. This means that the curve depends only on the cosmological parameters that govern evolution and not on initial conditions. Perturbations starting anywhere in phasespace asymptote to the Zeldovich curve as they evolve. \cite{nadkarni-ghosh_non-linear_2013} also showed that this invariance can be potentially exploited to break parameter degeneracy between parameters that govern initial conditions such as $\sigma_8$ or $n_s$ and those that govern evolution such as $\Omega_m$ or the equation of state $w$. In this paper, we evolve smoothed, compensated spherical top-hat overdensities and examine the resulting density and velocity profiles using a similar phase space description. Is there a unique relation between the density and velocity divergence akin to GR that is invariant of the initial conditions ? For simplicity, we focus on the $f(R)$ model and in particular consider the form given by \cite{hu_models_2007}. 

Spherical collapse (SC) in the context of MG has been considered by various authors. \citet{schafer_spherical_2008} used the Birkoff's theorem to write equations of motions for the outer radius of a top-hat for a phenomenological model that interpolated between $\Lambda$CDM and the DGP model. They argued that although Birkoff's theorem is known to be violated in most modified gravity models \citep{dai_consequences_2008}, the error associated with this approximation was small. \cite{schmidt_spherical_2009} coupled SC with the halo model to compute the halo mass function, halo bias and the non-linear matter power spectrum for the DGP theory. \cite{chakrabarti_spherically_2016} investigated collapse in $f(R)$ theories with a perfect fluid to understand the nature of the singularity. In standard GR, SC is most commonly used to compute the critical overdensity for collapse $\delta_c$, an important ingredient of excursion set based approaches such as the Press-Schecter formalism and its extensions \citep*{press_formation_1974,Sheth_2001,paranjape_excursion_2013}. The same is true in MG models. \citet{Herrera_2017} computed $\delta_c$ for $f(R)$, whereas Li and collaborators gave excursion estimates of the mass function in Galileon and chameleon models and also investigated the effect of the environment \citep{li_excursion_2012,li_extended_2012,barreira_spherical_2013}. \cite{martino_2009} computed cluster counts in MG for a Yukawa-like potential. Very recently, SC has also been combined with Large Deviations Theory to make analytic predictions of the matter density probability distribution function \citep{cataneo2021matter}. Many of the above investigations model SC in modified gravity similar to that in standard GR, but with a modified effective Newton's constant. 

The treatment of spherical collapse in this paper is akin to that of \citet{borisov_spherical_2012} and \citet{kopp_spherical_2013} who solve  the coupled continuity, Euler, Poisson and scalar field equations iteratively for an $f(R)$ model.  \citet{borisov_spherical_2012}  solve it in the Lagrangian frame by computing the acceleration of spherical shells at each step, whereas, \citet{kopp_spherical_2013} solve the system directly in the Eulerian frame as a coupled non-linear PDE using realistic profiles obtained from peaks theory \citep{bardeen_statistics_1986}.  However, our treatment differs from these papers in some important aspects. Firstly, we assume that the Compton wavelength is purely time-dependent and does not depend on the density of the perturbation. With this approximation,  we cannot model the  chameleon screening mechanism wherein the Compton wavelength changes with the density  of the perturbation \citep{khoury_chameleon_2004}. However, we illustrate the effect of screening by considering different scales for the width of the top-hat and demonstrate the scale-dependent nature of non-linear evolution. Secondly, we assume that the perturbations obey GR at early epochs and switch to the $f(R)$ evolution at a relatively late time denoted as $a_{switch}$. This is necessary because the $f(R)$ model exhibits high frequency oscillations at early epochs \citep{song_large_2007,hu_models_2007} making the equations numerically stiff and computationally expensive. However, starting the evolution too late, destroys consistency with the background $\Lambda$CDM evolution. Thus, the switching epoch, needs to be chosen judiciously.  In this paper, we perform an eigenvalue analysis of the background equations to estimate the magnitude of oscillations and make an informed choice of $a_{switch}$. Such an analysis has not been presented in the literature thus far. Finally, we do not solve the equations for the scalar field, but instead, recast the new degree of freedom in terms of the variable $\chi$ which is proportional to the difference of the two perturbation potentials $\Phi$ and $\Psi$. We present a closed form analytic solution for $\chi$ in spherical symmetry. We solve the system using a hybrid Lagrangian-Eulerian iterative scheme, wherein, the spatial equations for the two potentials are solved analytically in the Eulerian frame and the continuity and Euler equation are solved in the Lagrangian frame by recasting them as second order evolution equations for the radii of a series of concentric shells.

The rest of the paper is organized as follows. In \S \ref{sec:backequations} we define the Hu-Sawicki system and perform the eigenvalue analysis of the same. \S \ref{sec:pert} lays down the set of equations and initial conditions governing the non-linear evolution. \S \ref{sec:method} outlines the iterative method of solution and gives analytic expressions for the potential $\chi$. In order to illustrate screening, we plot the static solutions to the potentials as a function of the radial coordinate for different choices of the perturbation scale. \S \ref{sec:linear} evolves the compensated top-hat profile in the linear regime and \S \ref{sec:non-linear} extends the evolution to the non-linear regime. The late time density and velocity fields are analysed by plotting them on the density-velocity divergence phase space and fitting formulae are provided in certain limits. We summarize in \S \ref{sec:conclusion}. The appendix \S \ref{app:tophatsetup} gives the details of the initial top-hat setup and \S \ref{app:error} gives the details of the error tests performed to validate the numerical code. 

\section{Background cosmology in the Hu-Sawicki model}
\label{sec:backequations}
\subsection{Equations of motion}
We consider a $f(R)$ model of modified gravity for which the action has the form 
\beq 
S = \frac{1}{2 \kappa^2} \int d^4x \sqrt{-g} \{R + f(R)\} + S_m, 
\eeq
where $\kappa^2= 8 \pi G$ and $S_m$ is a minimally coupled matter action. The field equations obtained by varying the action with respect to the metric $g_{\mu \nu}$ are 
\beq 
(1+f_R) R_{\mu \nu} - \frac{1}{2} g_{\mu \nu} (R + f) + (g_{\mu \nu}  \Box -  \nabla_\mu \nabla_\nu)f_R = \kappa^2 T_{\mu \nu}, 
\eeq
where $T_{\mu \nu}$ is the stress-energy tensor, $R_{\mu\nu}$ and $R$ are the metric-dependent Ricci tensor and scalar and $f_R = df/dR$. For the spatially flat FRW background metric given by $ds^2 = -dt^2 + a(t)^2 (dx^2 + dy^2 + dz^2)$,
\beq
 R = 12 H^2 + 6 H H',
\label{Req}
\eeq
where $H$ is the Hubble parameter and prime denotes derivative w.r.t. $\ln a$. Assuming that dark matter is well described by a perfect fluid with zero pressure, the stress-energy tensor is 
\beq 
T_{\mu \nu } = \rho U_\mu U_\nu, 
\eeq
where $U^\mu$ is the rest-frame four-velocity, $\rho$ is the energy density. The Friedman equation for $f(R)$ models is then given by 
\beq 
H^2  - f_R (H H' + H^2) + \frac{1}{6} f + H^2 f_{RR} R'  = \frac{\kappa^2 {\bar \rho}}{3}. 
\label{backeq1}
\eeq
In standard GR, the Friedmann equation is a second order equation for the scale factor $a$. In $f(R)$ gravity, replacing $R$ using \eqnref{Req} 
recasts the Friedmann equation as a fourth order equation in time for the two unknown functions $f$ and $a$. There are two approaches to solve \eqnref{backeq1}.  One can assume a particular functional form for $f(R)$ and \eqnref{backeq1} reduces to a second order equation for the scale factor akin to standard GR. Examples include the Starobinsky and Hu-Sawicki models \citep{starobinsky_new_1980, hu_models_2007}. Alternatively, one can adopt the `designer approach'  wherein one demands that the background evolution be identical to the $\Lambda$CDM evolution and solves the resulting equations for $f$ as a function of time \citep{song_large_2007,pogosian_pattern_2008}. In this paper, we consider the former approach and in particular the Hu-Sawicki model as a specific example for which $f(R)$ is given by 
\beq 
f(R) = -m^2 \frac{c_1(R/m^2)^n}{c_2(R/m^2)^n + 1}. 
\label{fofR}
\eeq
$m^2 = H_0^2 \Omega_{m,0}$ denotes the mass scale, where $H_0$ and $\Omega_{m,0}$ are the Hubble constant and matter density parameters today. There are three parameters in this model: $c_1, c_2$ and $n$.  Requiring that at high redshifts (large $R$) the $f(R)$ model should reduce to standard GR gives, 
\beq 
\frac{c_1}{c_2} \approx 6 \frac{\Omega_{\Lambda,0}}{\Omega_{m,0}}.
\eeq  
The parameters $n$ and  $c_1/c_2^2$ are related to the value of the derivative $f_{R0}$ through 
\beq 
f_{R0} \approx -n \frac{c_1}{c_2^2} \left(\frac{12}{\Omega_{m,0}} -9\right)^{-n-1}. 
\eeq
Thus, choosing $n$ and $f_{R0}$ completely specifies the model. Throughout this paper, we choose $n=1$. Any theory of modified gravity needs to satisfy certain viability conditions \citep{pogosian_pattern_2008}. They are: $f_{RR} >0$ to ensure a stable high-curvature regime, $1+f_R >0$ to have a positive effective Newton's constant, $f_R<0$ and must asymptote to zero for large $R$ so as to ensure GR at early epochs and finally, $f_{R0}$ should be small enough to satisfy solar system and galactic constraints. The last condition imposes the bound: $|f_{R0}| \lesssim 10^{-6}$ \citep{hu_models_2007}. 

We use the value $f_{R0} = -10^{-6}$ for most of this paper. However, in \S \ref{sec:consistency} and \S \ref{sec:eigenanalysis} we use $f_{R0} = -0.01$ because the new aspects of the analysis are best illustrated with a larger $f_{R0}$. For the $\Lambda$CDM model, 
\bea
\label{HGR}H_{\Lambda CDM}^2(a) &=& H_0^2\left(\frac{\Omega_{m,0}}{a^3} + \Omega_{\Lambda,0} \right)\\
{\rm and} \;\; R_{\Lambda CDM}(a) &=& 3H_0^2 \left(\frac{\Omega_{m,0}}{a^3} + 4 \Omega_{\Lambda,0}\right).
\label{RGR}  
\eea
\subsection{Consistency  with $\Lambda$CDM }
\label{sec:consistency}
The explicit time dependence of $f(a)$ can be obtained by substituting \eqnref{RGR} in \eqnref{fofR}. With this substitution, \eqnrefs{Req} and \eqnrefbare{backeq1} become a two dimensional coupled system for the variables $R$ and $H$. An important, but not obvious, assumption in $f(R)$ models is that the solutions for $R$ and $H$ obtained from solving  \eqnrefs{Req} and \eqnrefbare{backeq1} are `consistent' with the $\Lambda$CDM values, namely, the evolution given by  \eqnrefs{HGR} and \eqnrefbare{RGR}. We explicitly perform this `consistency check' to validate this assumption. Following \cite{hu_models_2007}, we define the variables 
\beq 
y_H = \frac{H^2}{m^2} - a^{-3} \;\;\; {\rm and} \;\;\; y_R = \frac{R}{m^2} - 3 a^{-3}. 
\eeq
These variables have the advantage that their values are constant for the $\Lambda$CDM model given by $y_{H,\Lambda CDM} = \Omega_{\Lambda,0}/\Omega_{m,0}$ and $y_{R,\Lambda CDM} = 12 \Omega_{\Lambda,0}/\Omega_{m,0}$.
In terms of these variables,  \eqnrefs{Req} and \eqnrefbare{backeq1}  become 
\bea
 \label{eqnyh} y_H'&=& \frac{1}{3} y_R - 4 y_H, \\
\label{eqnyr} y_R'&=& \frac{9}{a^3} -\frac{1}{{\tilde f}_{{\tilde R}{\tilde R}}(y_H + a^{-3})} \left[  y_H  - {\tilde f}_{\tilde R} \left(\frac{y_R}{6} -y_H -\frac{1}{2 a^3}\right) +\frac{\tilde f}{6}\right], 
\eea
where 
\beq 
{\tilde f} = \frac{f}{m^2} \;\;\; {\rm and} \;\;\; {\tilde R} = \frac{R}{m^2}.
\eeq
The choice of the initial epoch is arbitrary and GR is assumed to be valid at that time. The effective equation of state \footnote{$w_{eff}$ is defined through the relation $H^2 = H_0^2\left(\frac{\Omega_{m,0}}{a^3} + \Omega_{de,0} \exp\left[{-\int_{a}^1 \left\{1+ w_{eff}(u)\right\} \frac{du}{u}}\right]\right)$, $u$ being a dummy variable. } in terms of these variables is 
\beq 
1+ w_{eff}(z) = -\frac{1}{3} \frac{y_H'}{y_H}.
\label{eqnofstate}
\eeq 
\begin{figure}
\includegraphics[width=18cm]{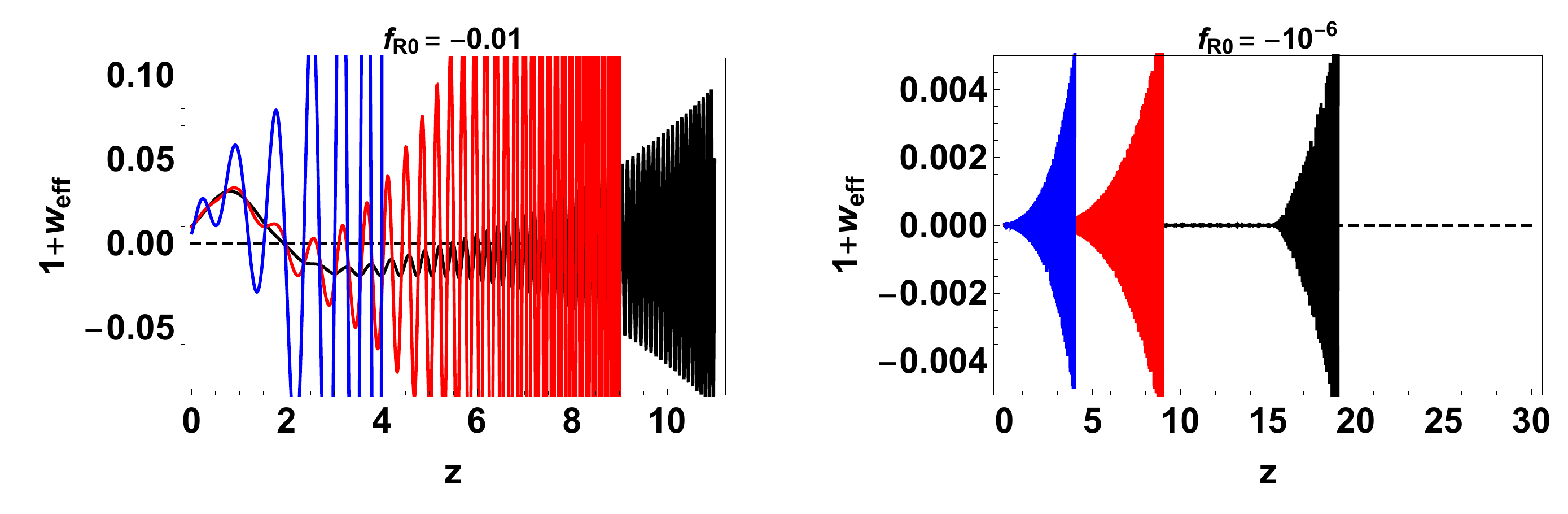}
\caption{ The deviation of the effective equation of state from its $\Lambda$CDM value as defined in \eqnref{eqnofstate} for $f_{R0} = -0.01$ (left panel) and $f_{R0} = -10^{-6}$ (right panel) plotted for three different starting epochs: $z_{init} = 4$ (blue), 9 (red) and 19 (black). The solutions are oscillatory. The amplitude of deviation is high at earlier epochs and decreases as the evolution proceeds. Thus, consistency with $\Lambda$CDM at low redshifts i.e., in the `dark energy era' is better if the evolution is started earlier. However, the oscillation frequency is higher when the evolution is started earlier making the system numerically `stiff'. The right panel shows that this deviation is significantly smaller for smaller values of $|f_{R0}|$, however the oscillation frequency is higher (see \figref{figeigen}).}
\label{weff_fig}
\end{figure}


Consistency with $\Lambda$CDM demands that $w_{eff}$ be close to -1. We find that $w_{eff}$ is oscillatory and in particular the deviation from the $\Lambda$CDM value depends sensitively on the starting epoch $a_{init}$.  
\capfigref{weff_fig} shows $w_{eff}$ for two values of $f_{R0}$ and three values of $a_{init}: 0.2, 0.1$ and  $0.05$ corresponding to $z=4$ (blue), 9 (red) and $19$ (black) respectively.  In order to compare with \cite{hu_models_2007}, we choose $\Omega_{m,0} = 0.24$ and $\Omega_{\Lambda,0} = 0.76$ for this analysis but the discussion is not sensitive to this choice. From the left panel, it is clear that the deviation from $\Lambda$CDM at lower redshifts is less 
when the system has a higher starting redshift. However, the oscillation frequency is higher at higher redshifts. Computationally, this means that the equations are more `stiff' requiring a higher number of steps when starting at a higher redshift. This behaviour poses a challenge: starting evolution too early is computationally expensive whereas starting too late implies greater deviation from the $\Lambda$CDM equation of state. This challenge is somewhat mitigated when smaller values of $|f_{R0}|$ are considered (right panel). The equations remain computationally stiff at high redshifts, however, the oscillation amplitude is significantly reduced implying an evolution very close to $\Lambda$CDM. For $a_{init} = 0.2$, the deviation of the equation of state, from its $\Lambda$CDM value, in the dark energy dominated era can be more than 5 \% when $f_{R0} = -0.01$ and is less than 0.05 \%  when $f_{R0} = -10^{-6}$. Thus, for parameter values of this model that satisfy solar system constraints, the oscillatory behaviour can be circumvented by starting the evolution later. In order to understand this behaviour better and make an informed choice about the starting redshift, we examine the above system by analysing the eigenvalues which give insights into the nature of oscillations.

\subsection{Eigenvalue analysis of the Hu-Sawicki background cosmology}
\label{sec:eigenanalysis}
\begin{figure}
\includegraphics[width=8cm]{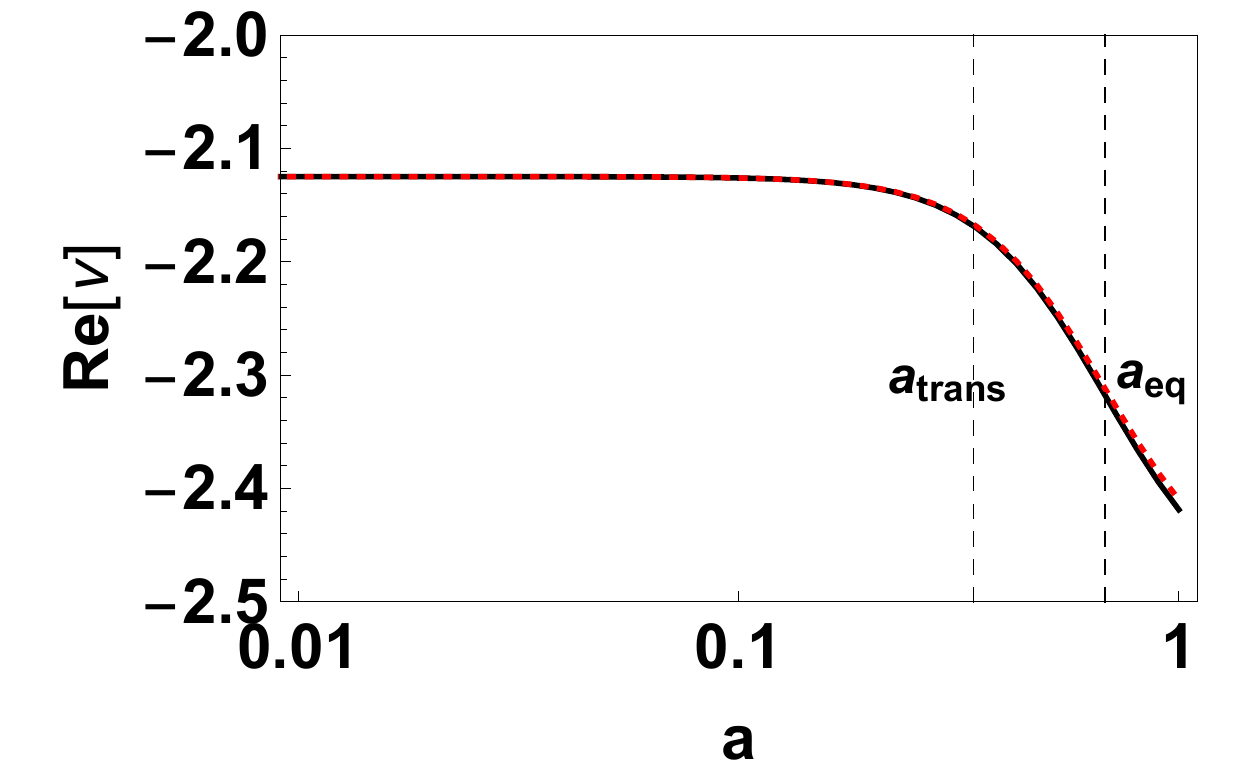}  \hspace{0.5cm}
\includegraphics[width=8cm]{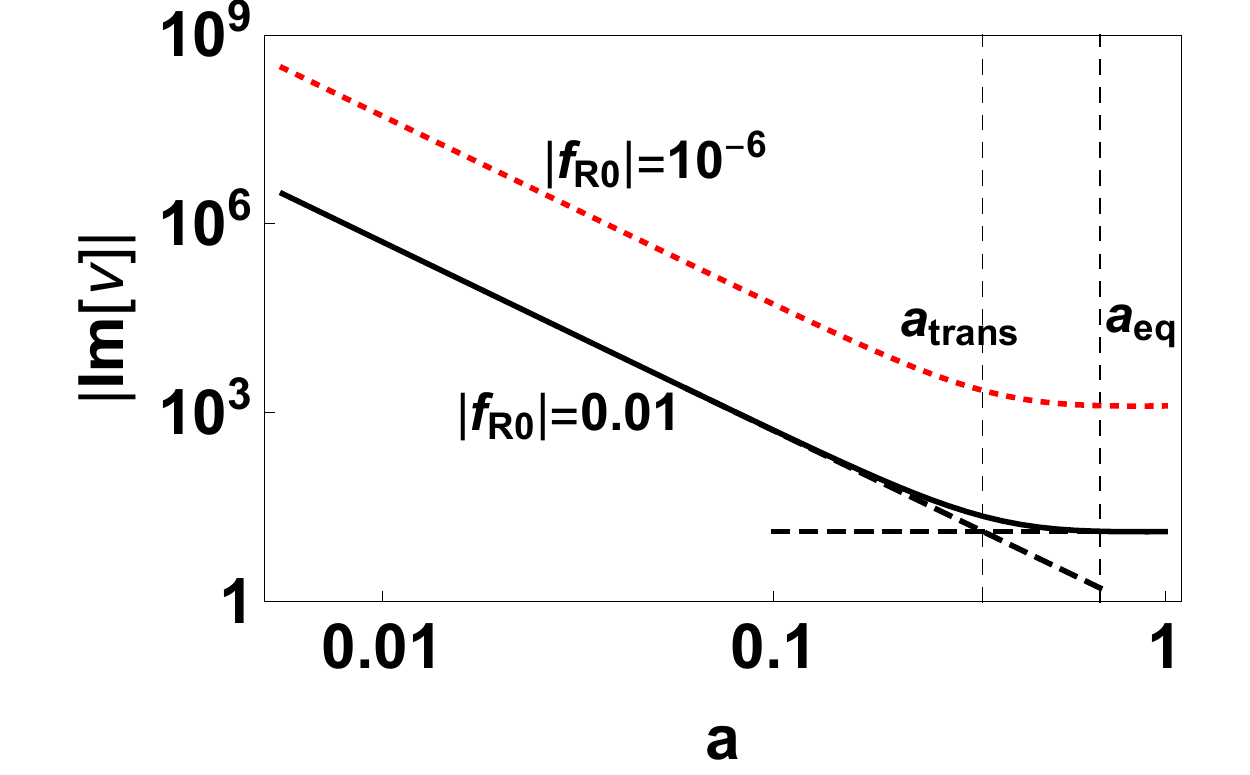}
\caption{Instantaneous eigenvalues of the linearized matrix ${\mathcal A}$ given by \eqnref{Amatrix}. The red dashed line indicates $f_{R0}=-10^{-6}$ and the black solid line indicates $f_{R0} = -0.01$; both models have $n=1$. The eigenvalues are complex implying oscillations. The real part (left panel) is negative implying that the oscillation amplitude decreases as $a$ increases. The imaginary part, which determines the oscillation frequency  is large at early epochs and decreases as a power law in $a$ and levels off as $a$ approaches 1. $a_{trans}$ denotes the epoch where the two asymptotes cross. At the same redshift, the instantaneous oscillation frequency is larger for smaller values of $|f_{R0}|$ but the transition value is independent of $f_{R0}$. The initial amplitude is lower for smaller values of $|f_{R0}|$ (not shown here), but the rate of decrease is independent of $|f_{R0}|$ as indicated by the real part. }
\label{figeigen}
\end{figure}

We recast the system given by \eqnrefs{eqnyh} and \eqnrefbare{eqnyr} in a more general form as
\beq 
\left\{y'_H, y'_R\right\} = \{ {\mathcal A}_1(y_H, y_R),  {\mathcal A}_2(y_H, y_R)\},  
\eeq
where ${\mathcal A}_1$ and ${\mathcal A}_2$ correspond to the right hand side of \eqnrefs{eqnyh} and \eqnrefbare{eqnyr}. If ${\mathcal A}_1$ and ${\mathcal A}_2$ were linear combinations of $y_H$ and $y_R$ with constant (in time) coefficients, then the system would be a linear, autonomous system. 
For such a system the eigenvalues of the operator matrix determine the nature of the solution. Real, negative eigenvalues indicate a stable solution, real positive eigenvalues indicate an unstable solution that grows with time and complex eigenvalues indicate oscillatory solutions. The rate of change of amplitude and frequency of the oscillations are determined by the real and complex part of the eigenvalues respectively. For a non-linear autonomous system, the eigenvalues of the linearized operator give information about the local behaviour of the solution. This interpretation of eigenvalues breaks down when the system is non-autonomous i.e., when the coefficients of the dependent variables are time varying \citep{ slotine,strogatz}. The system given by \eqnrefs{eqnyh} and \eqnrefbare{eqnyr} and  is both non-linear and non-autonomous. Thus, while the eigenvalues of the linearized system cannot give information about the global behaviour, they can still give qualitative information about the instantaneous solution.  In particular they can indicate the presence or absence of oscillations and provide an estimate of the instantaneous oscillation frequency. The linearized differential operator is given by 
\beq
{\mathcal A}_{lin} =\left( \begin{array}{cc}
\frac{\partial {\mathcal A}_1}{\partial y_H} &\frac{\partial {\mathcal A}_1}{\partial y_R}\\  \\
\frac{\partial {\mathcal A}_2}{\partial y_H}  &\frac{\partial {\mathcal A}_2}{\partial y_R}, 
\end{array} \right)
\label{Amatrix}
\eeq 
where 
\bea
\frac{\partial {\mathcal A}_1}{\partial y_H} &=& -4, \;\;\; \frac{\partial {\mathcal A}_1}{\partial y_R}=   \frac{1}{3},\\ 
 \frac{\partial {\mathcal A}_2}{\partial y_H} &=& -\frac {1 + {\tilde f}_{\tilde R}}{{\tilde f}_{{\tilde R}{\tilde R}} \left (a^{-3}+ y_H \right)} + \frac{\frac {{\tilde f}} {6} + y_H - {\tilde f} _ {\tilde R} \left (-\frac {1} {2 a^3} - y_H + \frac {y_R}{6} \right)} {{\tilde f} _ {{\tilde R} {\tilde R}}\left (a^{-3}+ y_H \right)^2} \;\;\;\;\;{\rm and} \;\;\;\; \;  \frac{\partial {\mathcal A}_2}{\partial y_R} = \frac{{\tilde f}_{\tilde R}}{6 {\tilde f}_{{\tilde R}{\tilde R}} \left(a^{-3}+y_H\right)}.
\eea
${\mathcal A}_{lin}$ can be evaluated at any time by using the values of ${\tilde f}, {\tilde f}_{\tilde R}, {\tilde f}_{{\tilde R} {\tilde R}}, y_H$ and $y_R$ at that time. The values of $y_H$ and $y_R$ can be read off from the numerical solution to \eqnrefs{eqnyh} and \eqnrefbare{eqnyr}. We found that using the GR values for $y_H$ and $y_R$ did not give significantly different answers to the eigenvalues.

\capfigref{figeigen} shows the eigenvalues of ${\mathcal A}_{lin}$.  The black solid line and red dashed lines correspond to $f_{R0} = -0.01$ and $f_{R0} = -10^{-6}$ respectively. 
The eigenvalues are complex. The left panel shows the real part of the eigenvalue, which is negative, indicating that the amplitude of the oscillations decreases as evolution proceeds. The right panel shows the imaginary part which indicates the instantaneous frequency of oscillations. The figure shows two distinct regimes. At very high redshifts, the oscillation frequency is large and decreases as a power-law in $a$:  $|\rm{Im}[\nu]| \sim a^\alpha$. The best fit value was found to be $\alpha \sim -3$.  At low redshifts, the instantaneous frequency is almost a constant. The epoch where these two asymptotes cross is marked as the transition epoch $a_{trans} \sim  0.34$ corresponding to a redshift of around $z \sim 2$. For a smaller value of $|f_{R0}|$, the magnitude of the imaginary part is higher implying higher frequency oscillations at the same redshift. However, we find that the transition epoch is independent of $|f_{R0}|$ and also independent of the value of $\Omega_{m,0}$ for flat models. 
For $a_{init}<<a_{trans}$, the high frequency oscillations make the system ``numerically stiff" and evolving the system is computationally expensive. 

The perturbed system described in the next section also exhibits oscillations akin to the background. In this paper, we have expressed the perturbation equations in terms of two temporal evolution equations for the density and velocity and two spatial equations for the metric potentials. This form of the equations is not amenable to the eigenvalue analysis discussed above. However, in order to overcome the issue of numerical stiffness we adopt the following strategy. We evolve the perturbations in standard GR starting from an early epoch $a_{init} = 0.001$ to a late epoch $a_{switch}$, after which the evolution `switches' over to $f(R)$ gravity. Following the insight gained from the eigenanalysis of the 
background we choose $a_{switch}$ to be in the vicinity of $a_{trans}$. In particular, we choose $a_{switch} = 0.1$. It is possible to choose $a_{switch}$ in a more quantitative manner by setting tolerances for deviation of equation of state or numerical stiffness, but we do not follow such a strategy here. We are interested in the qualitative differences in non-linear growth between the GR and $f(R)$ models. Furthermore, the DVDR, if such a unique relation exists, should depend only the evolution equations and not on the choice of $a_{switch}$. Thus, it suffices to only consider only a single choice for $a_{switch}$ and we do not investigate the dependence of non-linear growth on this parameter. Clearly, choosing $a_{switch}$ earlier (later) will increase (decrease) the effect of the modification. 

 High frequency oscillations in the Hu-Sawicki model have been referred to in the original paper \citep{hu_models_2007} as well as in follow up studies \citep{elizalde_oscillations_2012}. They have also been encountered in the context of evolution of designer models \citep{song_large_2007} as well as in linear perturbations (see for example \citealt{lima_linear_2013,pogosian_pattern_2008,silvestri_practical_2013}).  The issue of $\Lambda$CDM-consistency   has been explicitly addressed in the literature (for example \citealt{starobinsky_disappearing_2007,appleby_consistent_2007,ceron-hurtado_can_2016,chakraborty_model_2021}). Some of these studies have been in the context of finding stable $f(R)$ models. We note  that stability in the Hu-Sawicki model is guaranteed by the choice of parameters. The system is oscillatory but stable since the amplitude of oscillations decays as evolution proceeds. To the best of our knowledge, an understanding of the oscillatory behaviour based eigenvalue analysis of this system has not been presented so far in the literature. 
\subsection{Compton wavelength} 
In the Einstein frame, the $f(R)$ action can be recast as a standard Einstein-Hilbert action with non-minimally coupled scalar field corresponding to the additional degree of freedom represented by $f_R$ \citep{magnano_physical_1994,chiba_1r_2003,sotiriou_fr_2010}. The mass of this field is 
\beq
m^2_{f_R} = \frac{m^2}{3} \left(\frac{1 + {\tilde f}_{\tilde R}}{\tilde{f}_{{\tilde R}{\tilde R}}} - {\tilde R} \right).  
\eeq 
The associated physical Compton wavelength $\lambda_C$ \footnote{Putting in the dimensions this gives $\lambda_C =\frac{c \sqrt{3 {\tilde{f}_{{\tilde R}{\tilde R}}}}}{H_0 \sqrt{\Omega_{m,0}}}$.} is given by 
\beq 
\lambda_C = \frac{2 \pi}{m_{f_R}} \approx \frac{2 \pi \sqrt{3 {\tilde{f}_{{\tilde R}{\tilde R}}}}}{ m} . 
\label{Comptondef}
\eeq
We have used the fact that ${\tilde f}_{{\tilde R}{\tilde R}} <<1$ and ${\tilde f}_{{\tilde R}{\tilde R}}^{-1} >>{\tilde R}$ throughout the evolution. 
The comoving Compton wavelength $x_C$ and the reduced comoving Compton wavelength ${\bar x}_C$ are defined as 
\beq 
x_C = \frac{\lambda_C}{a}  \;\;\; {\rm and} \;\;\; {\bar x}_C = \frac{x_C}{ 2\pi} \approx   \frac{ \sqrt{3 {\tilde{f}_{{\tilde R}{\tilde R}}}}}{ a m},  
\label{Comptondef2}
\eeq 
where $m^2 = H_0^2 \Omega_{m0}$. 
This wavelength defines the `range' of the `fifth-force' corresponding to the additional degree of freedom represented by the scalar field. The effects of modified gravity manifest only on scales less than this range. The $f(R)$ model, obeys solar system constraints through the `Chameleon screening' mechanism, wherein the Compton wavelength decreases in high density regions reducing the range of the force \citep{khoury_chameleon_2004}.

\begin{figure}
\includegraphics[width=8cm]{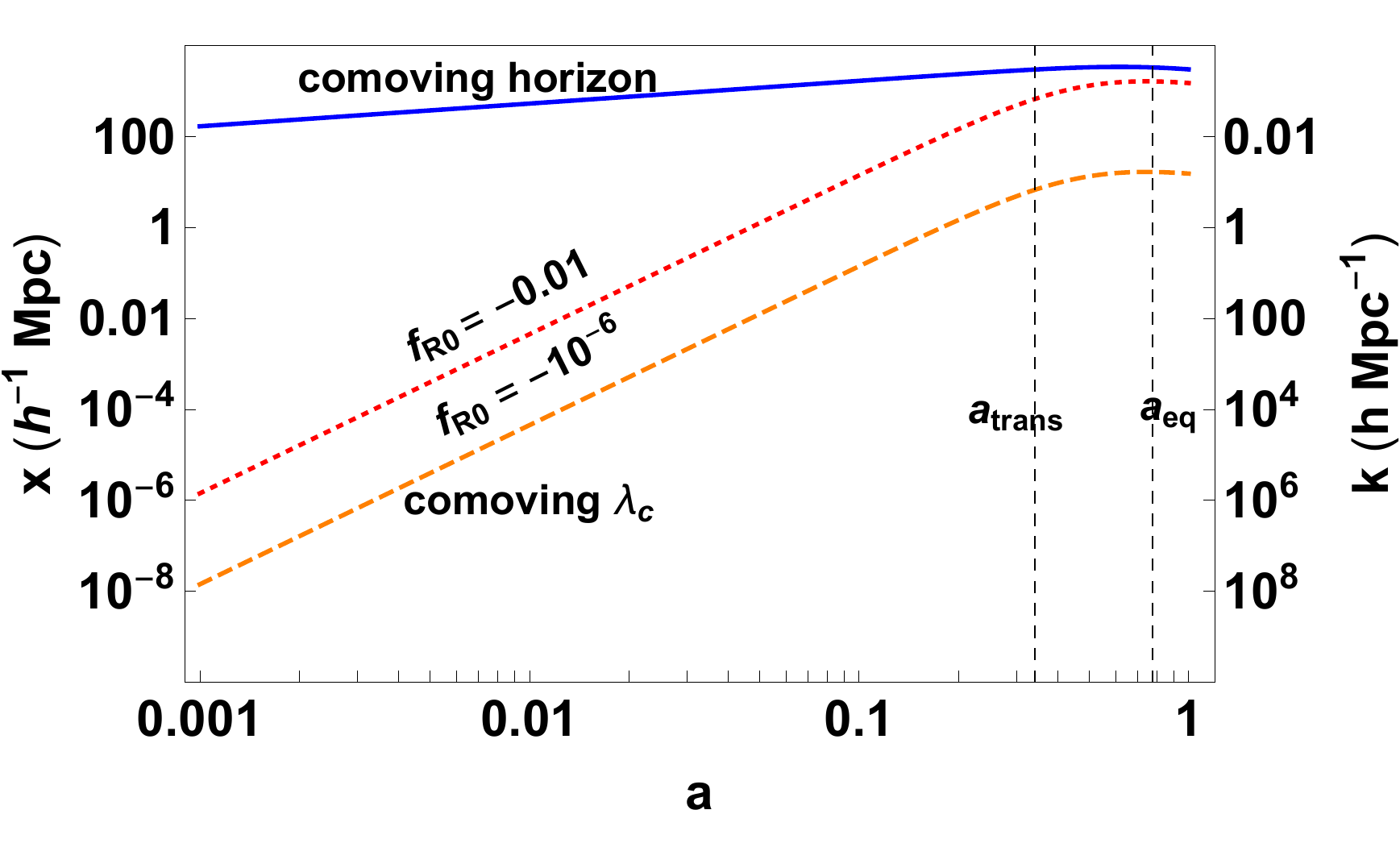} 
\caption{Important scales in the model marked on a $k-a$ plane (left panel). The red dotted and orange dashed lines denote the comoving Compton wavelength $x_C$ of the scalaron field implied by the Hu-Sawicki model for $f_{R0} = - 0.01$ and $f_{R0} = -10^{-6}$ respectively. At any epoch, the effects of modified gravity begin to manifest when $x \lesssim x_C$. The blue solid line indicates the scale when the given mode crosses the horizon. $a_{eq}$ denotes the epoch of dark energy-matter equality and $a_{trans}$ is the epoch when the frequency of oscillations in the Hu-Sawicki background start to level off to a constant. }
\label{fig:Compton}
\end{figure}

\capfigref{fig:Compton} shows the relevant scales and epochs on a plot in the $k-a$ plane. $a_{eq}$ corresponds to the epoch when the dark energy density is equal to the matter density in the background evolution. $a_{trans}$ denotes the epoch when frequency of oscillations in the Hu-Sawicki background starts to level off. Starting evolution for $a<<a_{trans}$ is computationally expensive where as $a>>a_{trans}$ implies significant deviations from the $\Lambda$CDM background, particularly for larger values of $|f_{R0}|$. $a_{trans}$ provides a estimate of the optimal epoch to start evolution. At early epochs, the cosmologically relevant scales ($\sim 1 h^{-1}$Mpc) are well above the Compton wavelength and one cannot expect to see departures from GR on those scales. Coincidentally, for $f_{R0}$ values which are allowed by solar system and galactic constraints the Compton wavelength becomes greater than $\sim 1 h^{-1}$Mpc at around $a \sim a_{trans}$.

\section{Equations for non-linear perturbations}
\label{sec:pert}
\subsection{Equations}
Assuming only scalar perturbations of the background, the metric in conformal Newtonian gauge is \citep{ma_cosmological_1995} 
\beq 
ds^2 = -a^2(1+2\Psi)  d\tau^2 + a^2(1-2 \Phi) (dx_1^2 + dx_2^2 +dx_3^2). 
\eeq 
$\Psi$ corresponds to the gravitational potential in the Newtonian limit which determines how particles move and $\Phi$ denotes the `curvature' fluctuation which is seeded by the matter density. ${\bf x} = (x_1,x_2,x_3)$ is the comoving coordinate and $d\tau = dt/a$ is the conformal time. 
The Universe is assumed to consist only of cold dark matter modelled to be a pressureless fluid with no anisotropic stresses. Let $\delta$ denote the fractional density and $v_i$ the coordinate velocity, which is equivalent to the peculiar velocity $a dx_i/dt$. We employ the Quasi Static Approximation (QSA) which comprises of two independent assumptions:  (1) the time derivatives of the potentials are small compared to the spatial derivatives and (2) the length scales are much smaller than the horizon scale, which is the usual `sub-horizon' approximation. In standard $\Lambda$CDM, these two assumptions are identical. The potentials vary on a time scale of $1/H$ and both assumptions correspond to $c k >> a H $, where $c$ is the speed of light. However, in modified theories, the potentials can vary on a time scale much smaller than the Hubble time. In particular, the potentials can be oscillatory, similar to the behaviour seen in the background. Suppose $\Psi \sim e^{i \omega \ln a} \Psi_0 $, where $\omega$ is the frequency of oscillation in the variable $\ln a$, then the first assumption reads $ck >>a \omega H $, which need not be satisfied even if the scales are sub-horizon since $\omega$ can be large. Thus, the two conditions need to be assumed separately. For the Hu-Sawicki model, with the parameters that satisfy solar system constraints, these oscillations are undetectable, justifying this assumption \citep{Hojjati_2012,silvestri_practical_2013}. 
In this limit, the Einstein equations are \citep{oyaizu_nonlinear_2008,schmidt_nonlinear_2009,borisov_spherical_2012}
\bea
\label{deltafReq}\nabla_x^2 \delta f_R &=& \frac{a^2 \delta R}{3c^2} -  \frac{H^2 a^2 \Omega_m \delta}{c^2}\\
\nabla_x^2 \Psi &=&  2 H^2 a^2 \Omega_m \delta - \frac{a^2 \delta R}{6}, 
\label{psieq}
\eea
where $\delta f_R$ and $\delta R$ are perturbations to $f_R$ and the Ricci scalar $R$ respectively. Note that $\delta R$ is an implicit non-linear function of $\delta f(R)$. However, assuming that the change in $R$ is small, Taylor expanding around the background gives 
\beq 
\delta f_R \approx f_{RR} \delta R,   
\label{lowcurv}
\eeq
where $f_{RR}$ is evaluated at the background $R$ given by \eqnref{RGR}. In the limit of $|f_R|<<1$ and $|f/R|<<1$, the difference between the two potentials $\Phi$ and $\Psi$ is related to $\delta f_R$ through (\citet{hu_models_2007})
\beq  
\Phi - \Psi =  c^2 \delta f_R.
\label{phipsi}
\eeq 
This relation is also follows from the anisotropy part of the Einstein equations in the QSA. It is further convenient to define the variables (\citealt{pogosian_pattern_2008})
 \bea
\label{chidef} \chi &=&  \Phi - \Psi \\
\label{Phidef} \Phi_+ & = & \frac{\Phi + \Psi}{2}. 
\eea 
Substituting \eqnrefs{lowcurv}, \eqnrefbare{phipsi}, \eqnrefbare{chidef} and \eqnrefbare{Phidef} in \eqnrefs{deltafReq} and \eqnrefbare{psieq} and combining with the conservation equations, which stay unchanged in $f(R)$, gives the system 
\bea 
\label{conteq}\frac{\partial \delta}{\partial t} + \left(\frac{{\bf v}}{a} \cdot \nabla_x\right) \delta &=& -\frac{(1+\delta)}{a} (\nabla_x \cdot {\bf v})\\
\label{euler}\frac{\partial {\bf v}}{\partial t} + \left(\frac{{\bf v}}{a} \cdot \nabla_x\right) {\bf v} +  H {\bf v}& =& -\frac{1}{a} \nabla_x \Psi\\
\label{phiplus}\nabla_x^2 \Phi_+& =&\frac{3}{2}  H^2  a^2 \Omega_m \delta \\
\label{chi}\nabla_x^2 \chi  - \frac{ a^2}{3 c^2 f_{RR}}  \chi &=&- H^2  a^2 \Omega_m \delta.
\eea
Note that $\Psi = \left(\Phi_+ - \frac{\chi}{2}\right)$. We have retained $\Psi$ in the right hand side of \eqnref{euler} to illustrate that the form of \eqnrefs{conteq} and \eqnrefbare{euler} is unchanged from the standard GR case. $ \Omega_m$ refers to the time-dependent matter density parameter, which is related to its value today as 
\beq 
\Omega_m(a) = \frac{H_0^2 \Omega_{m,0} a_0^3}{H^2 a^3}. 
\eeq
This is a coupled system of equations for the variables $\delta$, ${\bf v}$, $\Phi_+$ and $\chi$ which are functions of space as well as time. These have to be solved given the initial profiles for $\delta$ and ${\bf v}$ and boundary conditions for  $\Phi_+$ and $\chi$. Assuming no sources at infinity (homogeneous on large enough scales) and no forces at the origin (solution is not singular at $x=0$), gives the boundary conditions for \eqnrefs{phiplus} and \eqnrefbare{chi}
 \beq
  \Phi_+(x\rightarrow \infty) = 0, \;\;\; \chi(x\rightarrow \infty) =0,  \;\;\; \left.\frac{\partial \Phi_+}
  {\partial x}\right|_{x=0} = 0 \;\;\; {\rm and} \;\;\;  \left.\frac{\partial \chi}{\partial x}\right|_{x=0} =0.
 \eeq
In what follows, we set  $\Omega_{m,0} = 0.32$ and $\Omega_{\Lambda,0} =0.68$ in accordance with \cite{planck_collaboration_planck_2020}\footnote{https://www.cosmos.esa.int/web/planck} and we use the terms `standard GR' and $\Lambda$CDM interchangeably when we refer to this model. The density and velocity profiles are known as a function of the radial coordinate at some initial time $a_{init}$. Throughout this work, we consider compensated top-hats as initial conditions. The details of the set-up are given in appendix \ref{app:tophatsetup}. The perturbation is characterised by a single scale, namely, the width of the top-hat $x_{top}$. We note that the coefficient of $\chi$ in \eqnref{chi} can be re-written as ${\bar x}_C^{-2}$ using \eqnref{Comptondef2} and \eqnref{chi} becomes 
\beq 
\nabla_x^2 \chi  - \frac{1}{{\bar x_C}^2} \chi =- H^2  a^2 \Omega_m \delta.
\label{chi2}
\eeq
The solution to $\chi$ depends on the ratio\footnote{The $Q$ defined here is different but related to the $Q$ defined in Fourier space in \cite{pogosian_pattern_2008}.}
\beq 
Q =\frac{{\bar x}_C}{x_{top}}. 
\eeq
When $Q>>1$, \eqnref{chi2} reduces to $\nabla^2 \chi   \approx - H^2 a^2 \Omega_m \delta$ and $\chi$ is proportional to $\Phi_+$. In this limit, the gravitational potential $\Psi$ is enhanced by a factor of $4/3$ as compared to its value in standard GR (denoted by $\Psi_{GR}$) i.e., $\Psi = 4/3 \Psi_{GR}$. When $Q<<1$, the solution for $\chi$ is given by $\chi = {\bar x}_C^2 H^2 a^2 \Omega_m \delta$. The  dimensionless value of the extra field $\chi/(H^2 x_{top}^2)$ is suppressed by a factor of $Q^2$ and the evolution is close to GR.  $Q<<1$ has been referred to as the `Compton condition' in \citet{hu_models_2007}. 

\begin{figure}
\includegraphics[width=8cm]{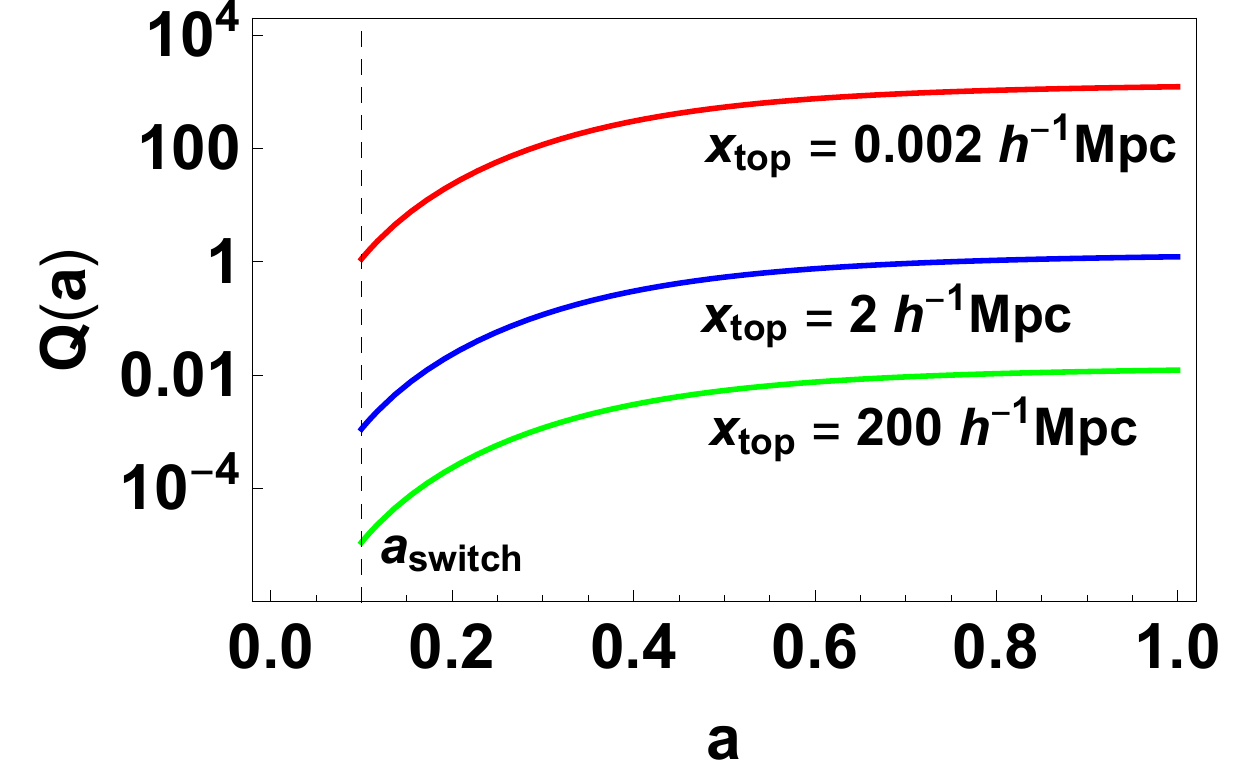}
\caption{The parameter $Q = {\bar x}_{C}/(x_{top})$ plotted as a function of $a$ for the three values of top-hat widths considered in this paper. The dotted line denotes $a_{switch}$, which is the epoch at which evolution is assumed to switch from GR to $f(R)$. In the text, we characterize these three fluctuations by their value of $Q$ today and ignore the time dependence. }
\label{fig:qvsa}
\end{figure}

In this work, we assume that ${\bar x_C}$ depends only on the value of $f_{RR}$ in the background and does not change with the perturbation. As a result, we do not capture the `chameleon screening mechanism' wherein the Compton wavelength is small in high density regions; this is responsible for satisfying solar system constraints \citep{khoury_chameleon_2004,hu_models_2007}. Nevertheless, we are able to illustrate `screening' by considering perturbations of differing length scales $x_{top}$ for a fixed $f_{R0}$. We consider three values of $x_{top} = 0.002, 2$ and 200 $h^{-1}$Mpc, which for $f_{R0} =  -10^{-6}$ gives $Q= 1217, 1.21, 0.012$ at $a=1$ respectively.
We refer to $Q>>1$ and $Q<<1$ as the `strong' and `weak' field regimes respectively. The evolution of $Q$ for these three scales is shown in \figref{fig:qvsa}. Note that, for $x_{top}  \sim 0.002 h^{-1} {\rm Mpc}$, $Q>1$ and for $x_{top}  \sim 200 h^{-1}$Mpc, $Q <1$ from $a = 0.1$ to today and they can be considered as being `strong' and `weak' perturbations over the entire domain of evolution. In what follows, we will refer to the three perturbations with their value of $Q$ today and ignore the temporal evolution of $Q$. 

\section{Method of solution}
\label{sec:method}
\subsection{Outline}
To proceed we further restrict ourselves to a spherically symmetric perturbation which is modelled as a series of concentric shells.  
\capeqnrefs{conteq} and \eqnrefbare{euler} involve time-derivative operators for the variables $\delta$ and ${\bf v}$, whereas  \eqnrefs{phiplus} and \eqnrefbare{chi} contain purely spatial derivatives. The gravitational potential depends on the instantaneous density making   
 \eqnrefs{conteq} to \eqnrefbare{chi} an integro-differential equation. To solve this system we employ an iterative scheme as follows. The system is evolved from the initial epoch, $a_{init} $, to the intermediate epoch, $a_{switch}$, in a single step using the equations of GR. The time domain from $a_{switch}$ to $a_{final}$ is divided into $N_t$ intervals. The density and velocity profiles are known at the beginning of each interval. We first solve the spatial differential equations \eqnrefs{phiplus} and \eqnrefbare{chi} in Eulerian space (fixed radial coordinates) to compute the gravitational potential corresponding to the density at the start of the step. Assuming that this potential is constant over the small time interval, we propagate the system using  \eqnrefs{conteq} and \eqnrefbare{euler} to obtain the density and velocity at the start of the next step. This temporal evolution is carried out in Lagrangian coordinates. The advantage of using Lagrangian coordinates is that the Lagrangian time derivative is the total time derivative 
 \beq 
\frac{d}{dt}  = \frac{\partial}{\partial t} + \frac{1}{a}({\bf v} \cdot \nabla_x).
\label{totalderivative} 
\eeq
This converts the non-linear coupled p.d.e. given by \eqnrefs{conteq} and \eqnrefbare{euler} to a second order ordinary differential equation. 
We perform error tests to check that this iterative hybrid Lagrangian-Eulerian scheme is convergent.   
\subsection{Lagrangian Coordinates}
 Let origin be the centre of the sphere. Define the Lagrangian coordinate of a shell to be its initial comoving coordinate, denoted by $q$ (we henceforth drop the vectors since the dynamics is only radial). The (Eulerian) comoving coordinate at any later epoch is 
\beq
 x\equiv x (q,a) = A(q, a) q, 
 \label{xdef}
\eeq
where $A(q,a)$ can be thought of as a `scale factor' of the shell at $q$. The $q$ dependence of the scale factor arises because, in general, the density is radially dependent. By definition, for all shells $A(q, a_{init}) = 1$. The peculiar velocity at any later time $t$ is 
\beq 
 v(q,a) = a {\dot A}(q,a) q = a H(a) A'(q, a) q, 
\label{Lagvel}
\eeq
where `dot' denotes the total derivative w.r.t. time $t$ and `prime' denotes derivatives w.r.t. $\ln a$. If $\delta(q, a_{init})$ is the density at the initial epoch, mass conservation for each shell implies that the density at any later epoch is 
\beq 
1+\delta(q, a) = \left (1+ \delta(q,a_{init})\right) \frac{q^2 dq}{x^2dx} =  \frac{(1+ \delta_{init})}{A^3 \left|1+ \frac{q}{A} \frac{dA}{dq}\right|}. 
\label{Lagdensity} 
\eeq
Using the definitions in \eqnrefs{xdef} and \eqnrefbare{Lagvel} and changing the derivatives to Lagrangian coordinates, it can be easily checked that the density given by the above expression satisfies the continuity equation given by \eqnref{conteq}. The spherically averaged density is defined as 
\beq
\Delta(x) = \frac{3}{x^3} \int_0^x  \delta(x,a) x^2 dx. 
\label{Deltadef}
\eeq
It satisfies the condition 
\beq 
1  + \Delta(q,a) = \frac{(1 +\Delta_{init}) A^3_{init}}{A^3}.
\label{Deltaconserve}
\eeq
\subsection{Equations in the hybrid Eulerian-Lagrangian scheme}
Substituting \eqnref{totalderivative} and \eqnref{Lagvel} in \eqnref{euler}, imposing spherical symmetry in \eqnrefs{phiplus} and \eqnrefbare{chi}, changing the time variable to $\ln a$ and defining ${\tilde \Phi} = \Phi/H^2, {\tilde \chi} = \chi/H^2$ and ${\tilde \Psi} = \Psi/H^2$ recasts the system as
\bea 
\label{eqnforA} A'' + \left(2 - \frac{3}{2} \Omega_m(a) \right) A' &=& -\frac{1}{a^2 q} \nabla_q {\tilde \Psi}(q,a) \\
\label{phipluseq}\frac{\partial^2 {\tilde \Phi_+}}{\partial x^2} + \frac{2}{x} \frac{\partial{\tilde \Phi}_+}{\partial x} &=& \frac{3}{2} a^2 \Omega_m(a) \delta(x,a). \\
\label{chieq}\frac{\partial^2 {\tilde \chi}}{\partial x^2} + \frac{2}{x} \frac{\partial{\tilde \chi}}{\partial x} - \frac{1}{{\bar x}_C^2} {\tilde \chi} &=& - a^2 \Omega_m(a) \delta(x,a),  
\eea 
where the `prime' denotes derivative with respect to $\ln a$. The boundary conditions for each step become
\beq
  {\tilde \Phi_+}(x\rightarrow \infty) = 0, \;\;\; {\tilde \chi}(x\rightarrow \infty) =0, \;\;\; \left.\frac{\partial {\tilde \Phi_+}}{\partial x}\right|_{x=0} = 0, \;\;\; {\rm and} \;\;\; 
  \left.\frac{\partial {\tilde \chi}}{\partial x}\right|_{x=0} =0.
  \label{boundarycond}
 \eeq
From \eqnref{Lagvel} and the definition of the Lagrangian coordinate, initial conditions for $A$ are: 
\beq
 A(q,a_{init}) = 1\;\;\; {\rm and}  \;\;\; A'(q, a_{init}) = \frac{v_{init}(q, a_{init})}{a_{init}H(a_{init}) q}.
 \eeq
In the GR regime, ${\tilde \Phi_+} = {\tilde \Psi}$. Using the solution to Poisson's equation in terms of $\Delta$, and the equations reduce to a second order ODE for $A$:
\beq
\label{GReqnforA} A'' + \left(2 - \frac{3}{2} \Omega_m(a) \right) A' = -\frac{\Omega_m(a) A}{2}  \left(\frac{1+ \Delta_{init}}{A^3} -1\right), 
\eeq
where $\Delta_{init}\equiv \Delta(q, a_{init})$ is the average density defined in \eqnref{Deltadef}, evaluated at the initial epoch. 
\subsection{Initial conditions} 
We evolve the system assuming GR from $a_{init} = 0.001$ to the switching epoch $a_{switch} = 0.1$. The initial profile for $\delta(x,a=0.001)$ is a smooth compensated top-hat. Its amplitude is chosen small enough so that the profile does not collapse in either $f(R)$ or GR but large enough to become non-linear. The details of the profiles used in various sections of the paper are given in appendix \ref{app:tophatsetup}. The velocity profile at $a=0.001$ is always set by assuming the Zeldovich condition \citep{zeldovich_gravitational_1970,brenier_reconstruction_2003}. For a radially varying profile, this gives 
\beq
v(x, a_{ainit}) = -\frac{f(\Omega_m) H a \Delta(x, a_{init}) x}{3},
\label{vinit}
\eeq 
where $\Delta(x,a)$ is defined in \eqnref{Deltadef} and $f(\Omega_m)\sim \Omega_m^{0.55}$ is the linear growth factor in the GR regime. For $a_{init} = 0.001$, $\Omega_m  \approx 1$. In 
Lagrangian coordinates, at the initial time, by definition, $x=q$ and the initial density and velocity are $\Delta(x,a_{init}) = \Delta(q, a_{init})$ and $v(x, a_{init}) = v(q, a_{init})$. Using \eqnrefs{xdef} and \eqnrefbare{Lagvel} gives the initial conditions 
\beq
A(q, a_{init}) = 1 \;\;\; {\rm for} \; {\rm all} \;\; q \;\;\;\;\; {\rm and} \;\;\; A'(q,a_{init})  = - \frac{\Delta(q, a_{init})}{3}. 
\label{GRinitalcond}
\eeq

\subsection{Algorithm} 
\label{sec:algorithm}
{\it Notation}:
Let $a_{init}$ and $a_{final}$ denote the initial and final epochs of interest and $a_{switch}$ denote the epoch where the evolution switches from GR to $f(R)$. The time between $a_{switch}$, to $a_{final}$ is divided into $N_t$ intervals equi-spaced in $\ln a$. Let $a_0 = a_{init}$, $a_1=a_{switch}$, $a_{N_t} = a_{final}$ and $a_2, a_3 \dots$ denote intermediate epochs. At any epoch $a_n$, the data at the start of the time interval is defined on a 1-d grid with $N_s$ points uniformly spaced in $\ln x$ space. These correspond to radial shells. $x_i(a_n), i = 1,2, \dots N_s$ denotes the Eulerian position of the $i$-th shell at the start of the $n$-th step. At $a_n$, the Lagrangian coordinate of the shell $i$-th shell is defined as $q_{n,i} = x_i(a_n)$. The density and velocity are known at the time $a_n$ on the uniform grid and are given by $\delta_i(q_n,a_n)$ and $v_i(q_n,a_n)$. Here we drop the subscript `i' on $q_n$ but it is understood that $q_n$ is different for each `i'. 

At the end of the step, i.e. at $a_{n+1}$, the Eulerian position, density and velocity of the $i$-th shell are denoted as $x_i(q_n, a_{n+1})$, $\delta_i(q_n, a_{n+1})$ and $v_i(q_n,a_{n+1} )$. However, the Eulerian positions $x_i(q_n, a_{n+1}), i = 1, 2, \dots N_s$ are not equi-spaced. At $a_{n+1}$, a new grid is defined in Eulerian space which is equispaced in $\ln x$ over the range $x_{min}$ to $x_{max}$, where $x_{min}$ and $x_{max}$ correspond to the Eulerian location of the innermost and outermost shell at $a_{n+1}$. 
This uniform grid is denoted as $x_{i'}(a_{n+1}), i' = 1, 2 \dots N_s$. The $i'$ indicates that the shell locations are re-defined after each step. Thus, $x_i(q_n, a_{n+1}) \neq x_{i'}(a_{n+1})$. The number of shells are the same at each step since no shells have collapsed or crossed. The Lagrangian coordinate at $a_{n+1}$ is given as $q_{n+1,i'} =  x_{i'}(a_{n+1})$. The density and peculiar velocity are obtained by interpolating $\delta_i(q_n, a_{n+1})$ and $v_i(q_n,a_{n+1} )$ onto the uniform grid. They are denoted as $\delta_{i'}(q_{n+1}, a_{n+1})$ and $v_{i'}(q_{n+1}, a_{n+1})$ respectively.  \\
Step 1. GR evolution:  
\begin{enumerate}
\item Evolve the system from $a_0 = a_{init}$ to $a_1 = a_{switch}$ using \eqnref{GReqnforA} and initial conditions \eqnref{GRinitalcond}. 
\item Construct the Eulerian position, density and velocity at $a_1$ using \eqnrefs{Lagvel} and \eqnrefbare{Lagdensity}. This gives $x_i(q_0,a_1)$, $\delta_i(q_0, a_1)$ and $v_i(q_0, a_1)$ for each shell $i$, $i = 1 \dots N_s$. 
\item Re-initialize the system by defining a uniform grid in $\ln x$ space, define the new Lagrangian coordinate $q_1$ and interpolate the density and peculiar velocity to get $\delta_{i'}(q_1, a_1)$ and $v_{i'}(q_1, a_1)$. 
\end{enumerate}
Step 2. $f(R)$ evolution: 
\begin{enumerate}
\item At $a_1$, and at any $a_n$ thereafter, first solve the spatial equations, \eqnrefs{phipluseq} and \eqnrefbare{chieq}, setting $a= a_n$ and $x = q_n$. The solutions for $\Phi_+$ and $\chi$ can be obtained analytically and are given in the next section. These expressions are evaluated numerically using the values of $\delta_i(q_n,a_n)$ known at the start of the step $a= a_n$. 
\item Using the solutions for $\Phi_+$ and $\chi$ compute the r.h.s of \eqnref{eqnforA} i.e., $\nabla_q \Psi(q, a) = \nabla_q \Psi(q, a_n)$ at each shell $i$. We assume that throughout the interval $a_n$ to $a_{n+1}$,  $\nabla_q \Psi_i(q_n, a) = \nabla_q \Psi_i(q_n, a_n)$ i.e., the force is constant. 
\item Compute the temporal evolution between $a_n$ and $a_{n+1}$ using \eqnref{eqnforA}. There are $N_s$ such equations, one for each shell. Let $A_i$ denote the scale factor of the $i$-th shell. The initial conditions are
\beq 
A_i(q_n,a_n) = 1 \;\;\; {\rm and} \;\;\;A_i'(q_n,a_n) =  \frac{v(q_n, a_n)}{a_nH(a_n) q_n} 
\eeq
\item Construct the density and velocity at $a_{n+1}$ using \eqnrefs{Lagvel} and \eqnrefbare{Lagdensity}. This gives $x_i(q_n, a_{n+1})$ and $\delta_i(q_n, a_{n+1})$ and $v_i(q_n, a_{n+1})$ for the $i$-th shell. Compute the end-points $x_{min}$ and $x_{max}$ at $a_{n+1}$ to be the Eulerian positions of the innermost and outermost shells. 
\item Re-initialize: define a new uniform grid over the range $x_{min}$ and $x_{max}$ denoted by  $x_{i'}(a_{n+1}), i = 1, 2 \dots N_s$. Define the new Lagrangian coordinate $q_{n+1} =  x_{i'}(a_{n+1})$ and interpolate the data given by $\delta_i(q_n, a_{n+1})$ and $v_i(q_n, a_{n+1})$  onto this uniform grid to get the initial data for the next step $\delta_{i'}(q_{n+1}, a_{n+1})$ and $v_{i'}(q_{n+1}, a_{n+1})$ at equi-spaced shell positions. 
\item Go to step 2(i) until $a_{n+1} =a_{final}$. 
\end{enumerate}

\subsection{Analytic Solutions for the potentials}
\label{sec:potentials}
The solutions to \eqnrefs{phipluseq} and \eqnrefbare{chieq} subject to the boundary conditions  \eqnref{boundarycond} are 
\bea
 {\tilde \Phi_+} (x) &=& -\frac{\mathcal P_1}{3} \int_x^{\infty} y \Delta(y) dy, \\
 {\tilde\chi}(x) &=& \frac{\mathcal P_3 }{\sqrt{\mathcal P_2}}\frac{1}{x} \left[ \exp(-\sqrt{\mathcal P_2}x) I_1(x) + \sinh(\sqrt{\mathcal P_2}x) I_2(x) \right]
 \eea
 where 
 \bea
 I_1(x) &=&  \int_0^x   \left\{ \sinh (\sqrt{\mathcal P_2}y) \delta(y) y \right\} dy,\\
I_2(x) &=& \int_x^\infty \left\{\exp(-\sqrt{\mathcal P_2}y)\delta(y)   y \right\} dy.
 \eea
 $y$ is a dummy variable in the integrals and $\Delta$ represents the spherically averaged fractional density contrast defined in \eqnref{Deltadef}.
${\mathcal P_1}$, ${\mathcal P_2}$ and ${\mathcal P_3}$ are purely time-dependent functions defined as
\beq
{\mathcal P}_1(a) = \frac{3 \Omega_m(a) a^2}{2}, \;\;\; {\mathcal P_2}(a) = \frac{1}{{\bar x}_C^2} \;\;\; {\rm and} \;\;\;  {\mathcal P_3}(a) = a^2 \Omega_m(a). 
\eeq
The `force' that moves the shells is related to the derivatives of the potentials: 
\bea
\label{gradphiplus}\nabla_x {\tilde \Phi_+} &=& \frac{\mathcal P_1}{3}x \Delta(x), \\
\label{gradchi}\nabla_x {\tilde \chi} &=& - \frac{{\tilde \chi} }{x}+ \frac{{\mathcal P_3}}{x} \left[-\exp(-\sqrt{\mathcal P_2}x) I_1(x) + \cosh(\sqrt{\mathcal P_2}x)  I_2(x) \right]. 
\eea
\begin{figure}
\includegraphics[width=18cm]{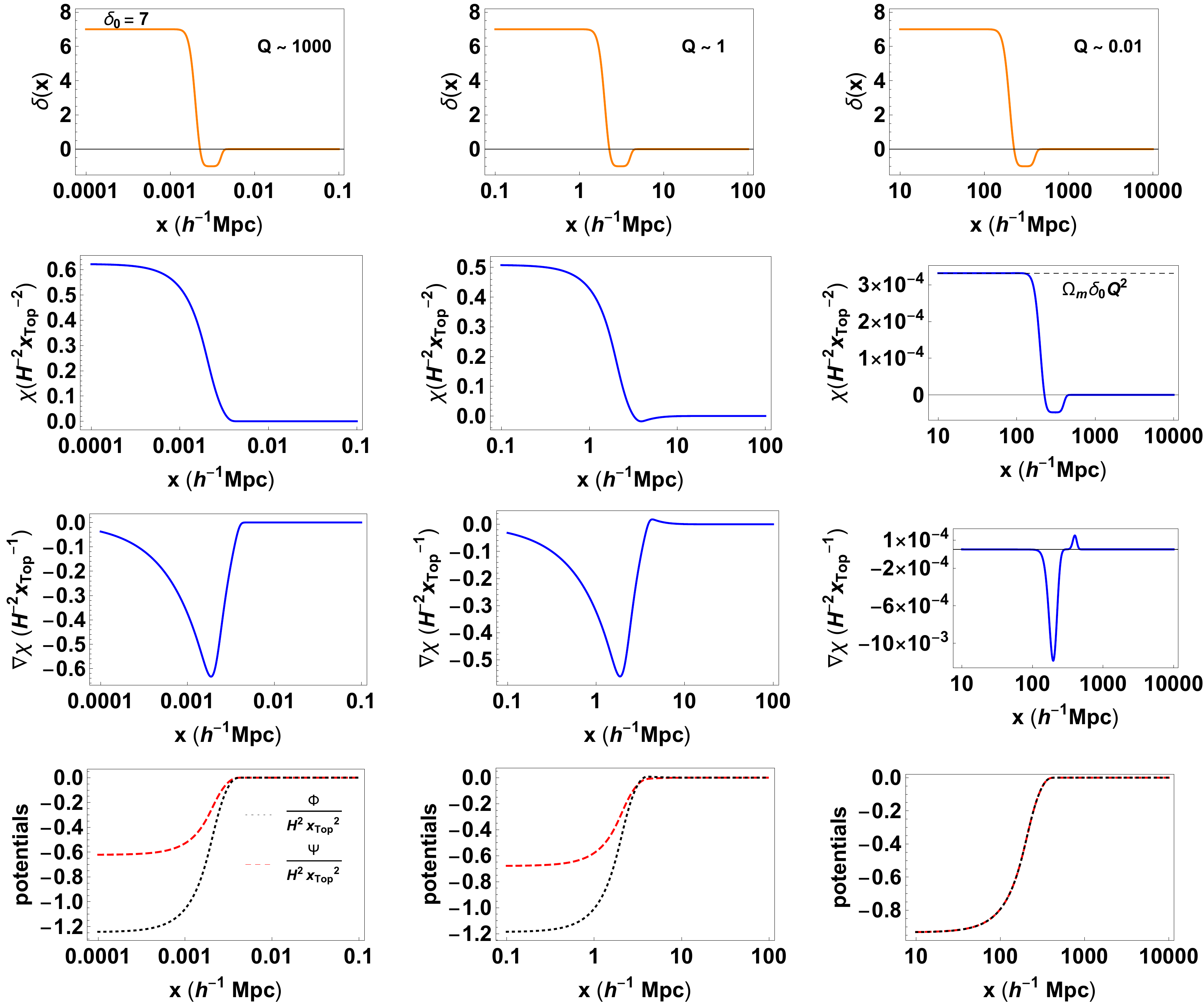}
\caption{The spatial solutions for dimensionless $\chi$, $\nabla \chi$, $\Phi$ and $\Psi$ at $a=1$ corresponding to a smoothed top-hat source $\delta(x)$  (top-panel). In the strong field limit $Q>>1$, the effect of the modification is the strongest corresponding to the largest difference between $\Phi$ and $\Psi$. In the weak field limit, $Q<<1$, the solution for $\chi$ is proportional to $\delta$ and the `extra force' which is proportional to $\nabla \chi$ is largest at the edge of the top-hat. The amplitude of this additional force is small compared to the potentials and they are effectively equal as is expected in standard GR. }
\label{staticsoln}
\end{figure}

\capfigref{staticsoln} shows the solution for the dimensionless $\chi$ (second panel), $\nabla \chi$ (third panel) and the potentials $\Phi$ and $\Psi$ (fourth panel) as a function of $x$ for the density profile $\delta$ corresponding to a smooth compensated top-hat (top-panel). The solution for $\chi$ is evaluated at $a=1$ for $f_{R0} = -10^{-6}$. The ${\bar x}_C$ for this $f_{R0}$ at $a=1$ is 2.43 $h^{-1}{\rm Mpc}$.  To illustrate how $\chi$ changes with the choice of scale we pick three values for the width of the top-hat denoted by $x_{top} = 0.002, 2$ and 200 $h^{-1}{\rm Mpc}$. This corresponds to $Q= 1217, 1.21, 0.012$ respectively. Changing the choice of scale is also equivalent to changing the value of $f_{R0}$. The effect of the fifth force depends on $\nabla \chi$. It is clear from the third panel, that when $Q<<1$, the extra force $\nabla \chi$ is non-zero only at the edge of the top-hat. In such a system, the evolved density develops a `spike' at the top-hat boundary (see \S \ref{sec:highNL}). This phenomenon has been reported in the literature before \citep{borisov_spherical_2012,kopp_spherical_2013,lombriser_parametrisation_2016}, but in the context of the chameleon mechanism. Here, we provide a mathematical  explanation of this feature based on the analytic solution, but in the absence of chameleon screening. A more detailed discussion can be found in \S \ref{sec:highNL}. The lowest panel shows the two potentials $\Phi$ and $\Psi$. When $Q>>1$, the two potentials clearly differ, when $Q<<1$, the effect of the `extra force' is effectively `screened' and the two potentials almost coincide as is expected in standard GR. The difference in the two potentials is related to $\chi$, which is suppressed by a factor of $Q^2$ in this regime. 

\begin{figure}
\includegraphics[width=8cm]{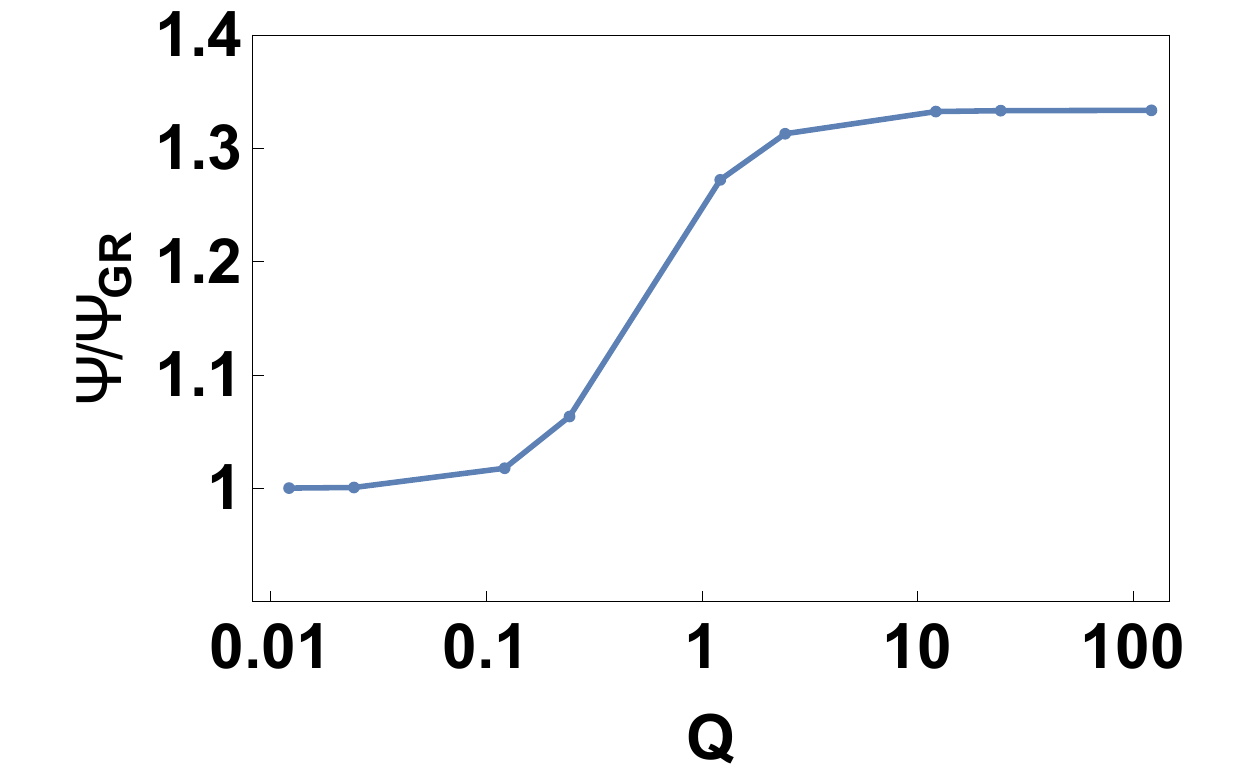}
\caption{The ratio $\Psi/\Psi_{GR}$ evaluated at the innermost shell for profiles with different values of $Q$. In $f(R)$, the `Newtonian potential' $\Psi$ i.e., the potential $\Psi$, that appears in Euler's equation is enhanced. In the strong field limit, $Q>>1$, the ratio $\Psi/\Psi_{GR} \sim 4/3$. In the weak field limit, $Q<<1$, the ratio asymptotes to unity.  }
\label{psiratio}
\end{figure}

\capeqnref{phipluseq} implies that the potential $\Phi_+$ has the same equation as the Newtonian potential in standard GR. We denote $\Phi_+ = \Psi_{GR}$. In the strong field limit, $\chi \approx -2/3 \Phi_+$ which gives $\Psi \approx 4/3 \Psi_{GR}$, whereas in the weak field limit $\Psi \approx \Psi_{GR}$. \capfigref{psiratio} shows the ratio $\Psi/\Psi_{GR}$, evaluated at the innermost shell, as a function of the parameter $Q$ of the profile. The strong and weak field asymptotic values are clearly in agreement with expectations. Since, the three values of $Q$ above were not sufficient to characterize this plot, the static potentials were evaluated for additional profiles with different values of $Q$. The details can be found in appendix \ref{app:tophatsetup}.

\section{Linear evolution} 
\label{sec:linear}
\begin{figure}
\includegraphics[width=18cm]{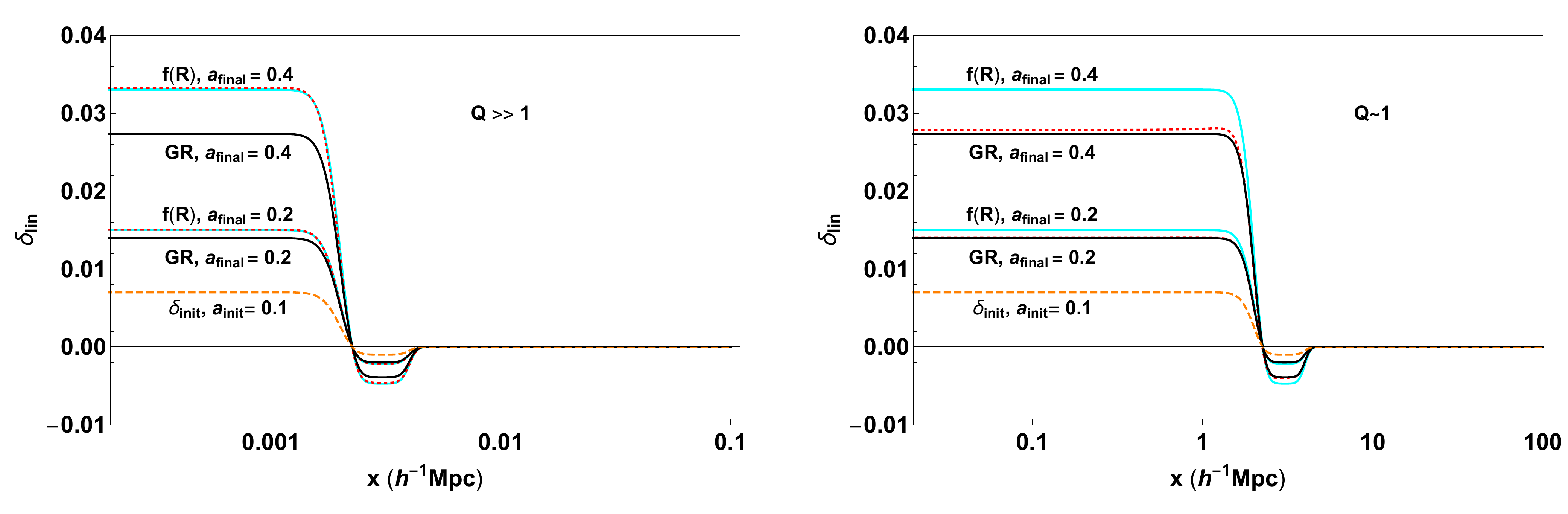}
\caption{Evolution in the linear regime. The left panel shows the initial density profile (orange, dashed), the expected final density in GR (black, solid), the theoretically expected (cyan) and the numerically (red, dotted) evolved final density in the $f(R)$ model at two final epochs. The cyan and red dotted curves coincide implying the correct linear evolution for $Q>>1$ given by the scale-independent \eqnref{deltalineqnMG}. The right panel shows the same data for a profile with $Q \sim 1$. The cyan and red-dotted curves deviate since the scale-independent linear growth equation is no longer valid. The deviation from GR is higher in the strong field regime ($Q>>1$) as compared to the intermediate field regime ($Q \sim 1$) as expected.  }
\label{lindelta}
\end{figure}
In GR, a top-hat remains a top-hat until shells cross. This follows as a consequence of the inverse-square law of Newtonian gravity. The evolution of each shell depends only on the mass enclosed inside the shell. The position dependence of the equations is implicit
for a radially varying density profile. This is not the case for $f(R)$ models. The force $\nabla_x \Psi$ in the Euler equation \eqnref{eqnforA} has a more complicated dependence on $x$ and this introduces a location-dependent evolution for each shell even for a top-hat density distribution. In linear theory, this manifests itself as a scale-dependent linear growth factor. However, as seen above, on scales much smaller than the Compton wavelength, $ \Psi \approx 4/3 \Psi_{GR}$ and the evolution is scale-independent. 
On these scales, the linear growth equation for $\delta$ which in standard GR is
\beq
{\rm Std. GR:} \;\;\; \;\;\;\;\;\; \delta'' + \left(2 + \frac{H'}{H}\right) \delta'  - \frac{3}{2} \Omega_m(a) \delta=0.
\label{deltalineqn}
\eeq
gets changed in modified gravity (MG) models to  
\beq
{\rm MG(Q>>1):} \;\;\; \delta_{MG}'' + \left(2 + \frac{H'}{H}\right) \delta_{MG}'  - 2 \Omega_m(a) \delta_{MG}=0.
\label{deltalineqnMG}
\eeq
The solution to this system can be written as 
\beq 
\delta_{GR}(a)  = \delta(a_{init}) \frac{D_{GR}(a)}{D_{GR}(a_{init})} = \delta(a_{init}) {\tilde D}_{GR}(a) \;\;\;\;{\rm and} \;\;\;\; \delta_{MG}(a)  =\delta(a_{init}) \frac{D_{MG}(a)}{D_{MG}(a_{init})} = \delta_{MG}(a_{init}) {\tilde D}_{MG}(a), 
\label{lineardelta}
\eeq
where $D_{GR}(a) $ corresponds to the growth factor given by the growing mode solution and we define the scaled growth factor ${\tilde D}(a)$ through the second equality. An analytic expression for $D(a)$ can be obtained for dark energy models \citep{dodelson}. 
To compute $D_{MG}$ for the $f(R)$ model, we solve \eqnrefs{deltalineqn} and \eqnrefbare{deltalineqnMG} with $\delta(a_{init})= 0.001$ from $a_{init} =0.1$ to $a_{final} =1$ with Zeldovich initial conditions ($\delta'(a_{init}) = \delta(a_{init})$). The ratio $\delta_{MG}(a)/\delta(a_{init})$ gives ${\tilde D}_{MG}(a)$.

To confirm the expectations in the linear regime, we evolve a compensated top-hat density from $a_{init} = 0.1$ up to two final epochs $a_{final} = 0.2$ and $a_{final} =0.4$ and for two values of $Q$:  $Q \sim 1217$ and $Q \sim 1.21$. The $Q<<1$ is not expected to show significant deviation from GR. Since the evolution is over a small time interval, it suffices to use a small number of steps: $N_t =40$. The initial velocity is set by assuming Zeldovich initial conditions. The initial and final densities are plotted in \figref{lindelta}. The dashed orange corresponds to the initial profile and the black line is the theoretically expected, linearly evolved profile in GR, the red-dotted line is the numerically evolved profile using the non-linear code for the Hu-Sawicki model with $f_{R0} = -10^{-6}$ and the cyan line is its theoretical expectation given by multiplying the profile with the scaled linear growth factor defined in \eqnref{lineardelta}. For the $Q>>1$, the cyan and the red-dotted lines coincide at both $a=0.2$ and $a=0.4$. Their deviation from the corresponding GR value  is larger for larger final epochs. This serves as a code check of the non-linear code in the linear regime. For $Q \sim 1$, the red-dotted lines and the cyan lines do not coincide because  \eqnref{deltalineqnMG} is no longer valid. The red-dotted lines are now closer to the GR values because the strength of the modification reduces as $Q$ decreases. It is important to note that the linear growth given by \eqnref{deltalineqnMG} is in real space. In Fourier space, it is possible to write a linear growth equation for each mode $\delta_k$ which is valid for all scales, but is scale-dependent (see for e.g., \citealt{bean_dynamics_2007, pogosian_pattern_2008}). Most current and upcoming large scale structure surveys aim to constrain modified gravity parameter by measuring the linear growth rate of perturbations. Most of these analyses are scale-independent i.e., they only look for a deviation from the growth rate in GR (for e.g. \citealt{alam_testing_2020}).  Although, a systematic scale-dependent measurement of the growth rate is ideal, scale-independent measurements can be used to place upper limits on the Compton wavelength of the scalaron field.

\section{Non-linear regime}
\label{sec:non-linear}
\begin{figure}
\includegraphics[width=16cm]{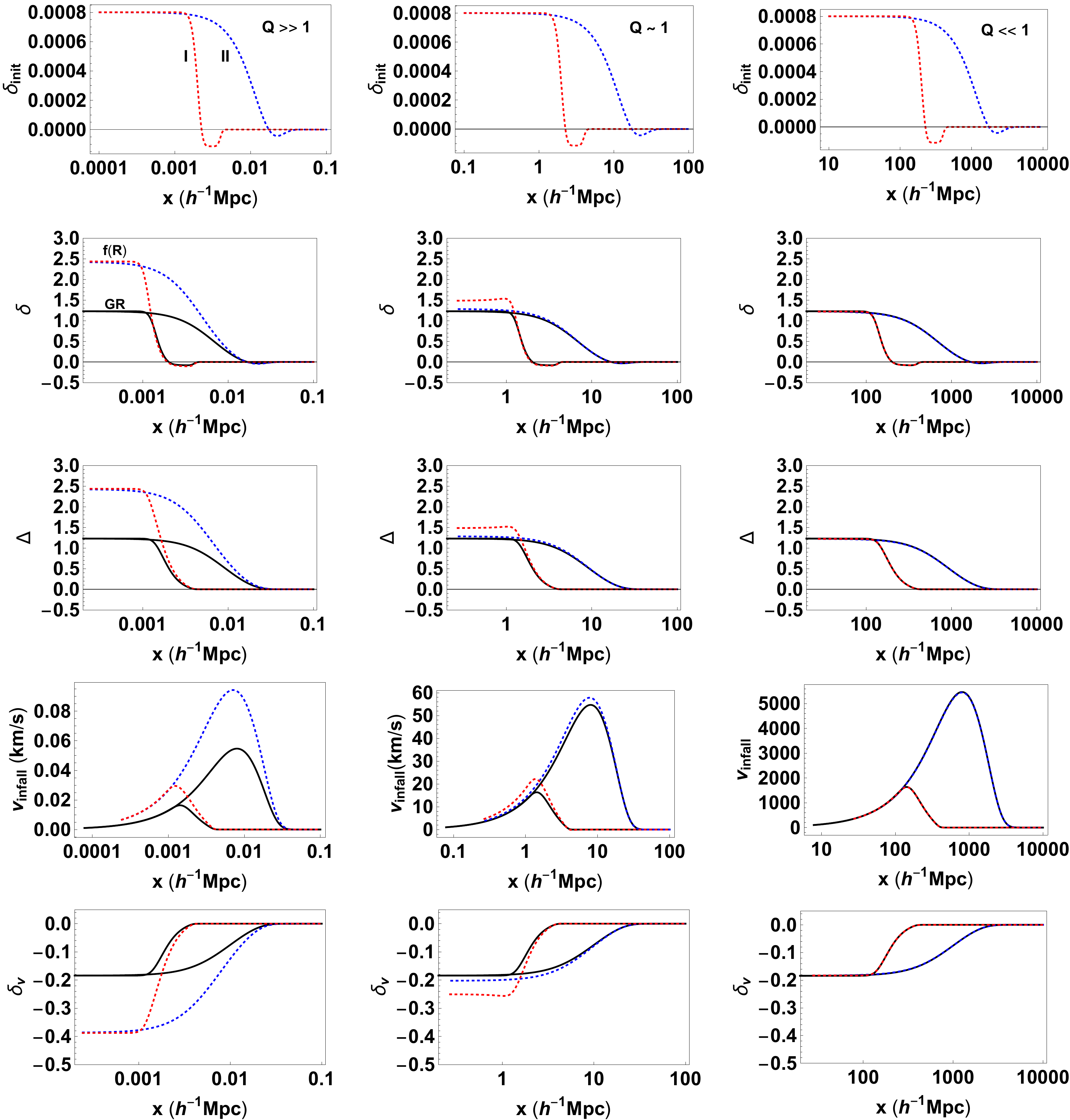}  
\caption{Initial $\delta$ at $a=0.001$ and the non-linearly evolved $\delta$, $\Delta$, infall velocity and relative Hubble velocity ($\delta_v$) at $a=1$ for three values of the top-hat scale $x_{top} = 0.002, 2, 200$ $h^{-1}{\rm Mpc}$ corresponding to $Q>>1$, $Q \sim1$ and $Q<<1$ respectively. 
The red and blue dotted curves correspond to $\sigma = 0.0025$ (profile I) and $\sigma = 0.1$ (profile II) evolved with the Hu-Sawicki model with $f_{R0} = -10^{-6}$. The solid curves denote the evolution using standard GR from $a=0.001$ to $a=1$ for the same. The evolution in GR is scale independent. Thus, the final profiles are the same for all values of $Q$. For the $f(R)$ model, When $Q>>1$, the evolution equations are structurally similar to GR with an enhanced potential. So the evolved profiles I and II coincide in the interior similar to the GR case, but with an increased final density. When $Q<<1$, the scale is outside the Compton scale making the system seemingly indistinguishable from GR.  When $Q \sim 1$, the evolution is scale-dependent and hence profile dependent. }
\label{denvel}
\end{figure}
\subsection{Non-linear density and velocity}

We evolve the compensated top-hat profile from $a_{init}=0.001$ to $a_{final}=1$ using the algorithm outlined in \S \ref{sec:algorithm} with $a_{switch} =0.1$. We consider three values of the top-hat scale $x_{top} = 0.002, 2, 200$ $h^{-1}{\rm Mpc}$ and two values of the smoothing parameter $\sigma = 0.0025$ (red, dotted, profile I) and $\sigma = 0.1$ (blue, dotted, profile II), for each scale.  The details of the initial profile are given in appendix \ref{app:tophatsetup}. \capfigref{denvel} shows the relevant quantities. The top-most panel shows the initial density profile which is the same for both models. The density and its spherical average are defined in \eqnref{Lagdensity} and \eqnrefbare{Deltadef}. The infall velocity and the relative Hubble velocity are defined as 
\bea
v_{infall} &=& - v(q,a)
\label{vinfalldef} \\
\delta_v &=& \frac{1}{H} \frac{\dot r}{r} - 1 = \frac{v(q,a)}{a H(a) x(q,a)} = \frac{A'}{A}. 
\label{deltavdef}
\eea
The black curves denote the evolution using standard GR from $a=0.001$ to $a=1$ and the dotted curves denote the evolution using the $f(R)$ model. 

In the $\Lambda$CDM model, the radial evolution of a shell depends only on the initial average density at the shell location and is independent of the size of the shell (see \eqnref{GReqnforA}). Consequently, in the GR case, the final density profiles for the two $\sigma$ values, coincide in the interior of the top-hat since the initial density in the interior is same for both profiles. Because of the scale-independent evolution, the final density profiles in GR are the  same for all values of $Q$. The magnitude of the velocity depends on the scale but the dimensionless quantity $\delta_v$ defined in \eqnref{deltavdef} is also independent of $Q$. 

In the $f(R)$ model, the evolution is, in general, scale-dependent. The evolution of the extra potential $\chi$ depends sensitively on the parameter $Q$. In the strong field regime when $Q>>1$, the structure of the equations is similar to that of GR. The final densities in the interior of the top-hat are independent of $\sigma$ as is the case in GR, but because the potential is enhanced by a factor of $4/3$, the final density of the top-hat is higher as compared to GR. 
In the intermediate regime, when $Q \sim 1$, the evolution is profile dependent. An un-smooth top-hat is described uniquely by a single scale $x_{top}$; smoothing broadens the top-hat edge increasing the effective width of the top-hat and decreasing the effective $Q$. Thus, the profile I (smaller $\sigma$) shows a greater density enhancement than profile II (larger $\sigma$). In the weak field regime, when $Q<<1$, the extra force is suppresed and the evolution is close to GR. There is expected to be some density enhancement at the edge of the top-hat as was discussed in \S \ref{sec:potentials} but it is too small to be seen on the scale of the plot. This effect is demonstrated in \S \ref{sec:non-linear}. 

\subsection{Phasespace evolution }
\label{sec:phasespace}
In standard gravity, the fractional overdensity and the divergence of the peculiar velocity are related through the coupled continuity and Euler equations. The temporal evolution of this system requires two initial conditions: the initial density and initial velocity field. However, it is physically reasonable to assume that there are `no perturbations at the big bang' and this condition relates the initial density and velocity uniquely. At first order, in Eulerian perturbation theory, this condition is implemented by ignoring the decaying mode in the linear solution for $\delta$ and in Lagrangian perturbation theory it is embedded in the `Zeldovich approximation'. This is usually referred to as the slaving condition since the velocity field is `slaved' to the acceleration field \citep{brenier_reconstruction_2003}. Thus, the linear density-velocity divergence (DVDR) relation is 
\beq 
\Theta = - f \delta, 
\eeq
where, $\Theta = \nabla_r \cdot {\bf v}/H$ is the scaled velocity divergence. The linear growth rate, $f$, is primarily dependent on the matter density $\Omega_m$ and is usually expressed as $f \equiv \Omega_m^\gamma$, where $\gamma$, sometimes called the growth index, is a sensitive probe of cosmology\footnote{$\gamma$ defined here differs from the post-Newtonian parameter sometimes used in literature \citep{bertschinger_distinguishing_2008, joyce_dark_2016}.}. For a pure matter universe $\gamma = 0.6$ \citep{Peebles80}, for a $\Lambda$CDM model $\gamma = 0.55$ \citep{linder_cosmic_2005} and for the DGP model of gravity, $\gamma = 0.68$ \citep{dvali_4d_2000,linder_parameterized_2007}.  

In \cite{nadkarni-ghosh_non-linear_2013}, N13, hereafter, this relation was extended to the non-linear regime using the spherical top-hat by imposing the condition `no perturbations at the big bang' in the full solution. 
At any epoch, given an initial $\delta$,  it is possible to compute the corresponding $\Theta$ that satisfies the condition using the exact evolution equation for the outer edge of the top-hat. This unique relationship traces out a curve in the two dimensional $\delta-\delta_v$ phasespace which was called the `Zeldovich curve' because it encapsulates the spirit of the Zeldovich approximation.  It is not obvious that this is the desired late time non-linear DVDR. In order to check this, N13, examined the dynamics of perturbations using the coupled continuity and Euler equations,  which for a top-hat in standard GR reduce to
\bea
\label{GRA1}\delta' &=& -(1+ \delta) \Theta \\
\Theta' &=& -\frac{3}{2}  \Omega_m(a) \delta - \Theta\left(2 + \frac{H'}{H}\right) -\frac{\Theta^2}{3}. 
\label{GRB1}
\eea
By evolving many initial $\delta-\Theta$ pairs using the above equations, it was shown that the `Zeldovich curve' is indeed an invariant of the dynamical system described by \eqnrefs{GRA1} and \eqnrefbare{GRB1}. This means that perturbations that start anywhere in phasespace, not necessarily on the curve, eventually converge to it. Those that start on the curve, stay on the curve. Thus, the curve traced out by the `no-perturbations at the big bang' condition indeed is the late time, non-linear DVDR. 

Finding such invariants of the dynamical system can be potentially useful in breaking parameter degeneracies as was illustrated in N13. This is because invariant sets depend only on cosmological parameters that appear as coefficients in the dynamical equations and are insensitive to the parameters that describe initial conditions, such as, the amplitude $\sigma_8$ or spectral index $n_s$. One of the primary aims of this paper is to understand if such invariants exist for the $f(R)$ system. Because of the complexity of the equations, it is not practically feasible to implement the algorithm used in N13 to select the $\delta-\Theta$ pairs that satisfy the `slaving condition'. Instead, we evolve the compensated top-hat profiles which start with Zeldovich initial conditions, compute the resulting density-velocity divergence pairs along the evolved curves and examine where they lie in phase space. Is the resulting curve in phase-space universal ? Or does it depend on the details of the initial profile ?

\begin{figure}
\includegraphics[width=18cm]{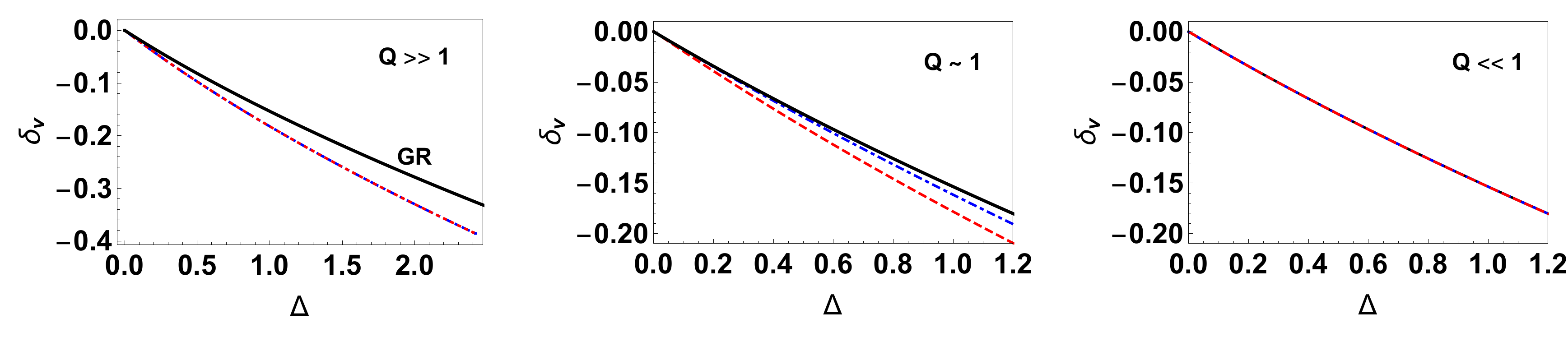}  
\caption{The spherically averaged density $\Delta$ and fractional Hubble parameter $\delta_v$ plotted on the $\Delta-\delta_v$ phase space for profiles I (red, dotted) and II (blue, dot-dashed). In standard GR, the `slaving condition' imposes a specific  relation between the non-linear density and velocity which traces out a special curve in the phasespace. This curve is an invariant of the dynamical system described by \eqnrefs{GRA2} and \eqnrefbare{GRB2}. In the $f(R)$ model, in the strong field regime ($Q>>1$), the evolution is scale-independent and the dynamics can be described by \eqnrefs{MGA2} and \eqnrefbare{MGB2} akin to standard GR and the non-linear relation between $\Delta$ and $\delta_v$ is profile independent. In the weak field limit ($Q<<1$), the curves coincide with the GR ($\Lambda$CDM) case.  However, in the intermediate regime ($Q \sim 1$), the Newtonian force is scale dependent and in the $\Delta-\delta_v$ plane, the dynamics is profile dependent.}
\label{phasespace}
\end{figure}

Since the density profile is radially varying, it is easier to use the spherically averaged $\Delta$ and the relative Hubble velocity defined in \eqnref{deltavdef} \footnote{N13, used the variables $\delta-\delta_v$ to describe the phase-space. For constant density perturbations considered in that paper, the two are equivalent $\Theta = 3\delta_v$.}.
In standard GR, using relations \eqnrefs{Deltaconserve},  \eqnrefbare{GReqnforA} and \eqnrefbare{deltavdef} the dynamical equations for the $\Delta-\delta_v$ system read: 
\bea
\label{GRA2}\Delta' &=& -3(1+ \Delta) \delta_v  \\
\label{GRB2}\delta_v' &=& -\frac{1}{2}  \Omega_m(a) \Delta -  \delta_v\left(2 + \frac{H'}{H}\right) -  \delta_v^2. 
\eea
The form of the coupled system given by \eqnrefs{GRA1}-\eqnrefbare{GRB1} or \eqnrefs{GRA2}-\eqnrefbare{GRB2} hinges on the fact that the potential in Euler's equation is related to the density through the Poisson's equation. For $f(R)$ theories, in general, this closure is not possible because of the extra degree of freedom encapsulated in $\chi$. However, in the limit $Q>>1$, $\Psi = 4/3 \Psi_{GR}$ and the system can be written as 
\bea
\label{MGA1}\delta_{MG}' &=& -(1+ \delta_{MG}) \Theta_{MG}, \\
\label{MGB1}\Theta_{MG}' &=& -2  \Omega_m(a) \delta_{MG} - \Theta_{MG}\left(2 + \frac{H'}{H}\right) -\frac{\Theta_{MG}^2}{3},
\eea
or alternately
\bea
\label{MGA2}\Delta_{MG}' &=& -3(1+ \Delta_{MG}) \delta_{v,MG},  \\
\label{MGB2}\delta_{v,MG}' &=& -\frac{2}{3}  \Omega_m(a) \Delta_{MG} -  \delta_{v,MG}\left(2 + \frac{H'}{H}\right) -  \delta_{v,MG}^2.
\eea
 \capfigref{phasespace} shows the $\Delta-\delta_v$ pairs plotted in the two dimensional phasespace for the profiles plotted in \figref{denvel}. These pairs are directly calculated from the evolved profiles at $a=1$ plotted in \figref{denvel} for all three values of $Q$. The red (blue) lines correspond to  profiles I and II with $\sigma = 0.0025$ and $0.1$ respectively. Note that, when $Q>>1$, although cases I and II have different profiles for $\Delta$ and $\delta_v$, the pairs, when plotted in phasespace, trace out the same curve. Thus, this curve is an invariant of the dynamics described by \eqnrefs{MGA2} and \eqnrefbare{MGB2}. Similarly, in the weak field limit, when $Q<<1$, the $\Delta-\delta_v$ pairs trace out the same curve for profiles I and II. In this case, this curve coincides with the GR limit and is an invariant of the dynamics given by \eqnrefs{GRA2} and \eqnrefbare{GRB2}. In the intermediate regime, when $Q \sim 1$, the $\Delta-\delta_v$ curve is different for profiles I and II. This is related to the fact that the dynamics in this regime cannot be described by a closed form dynamical system of the kind discussed above.

\begin{figure}
\includegraphics[width=18cm]{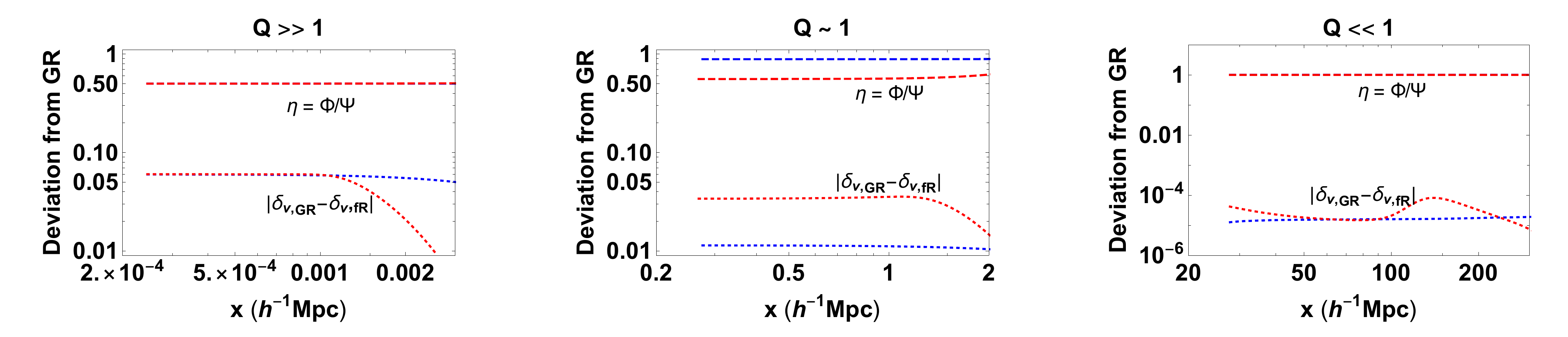}
\caption{Deviation of the DVDR relation in modified gravity from that in GR at $a=1$ corresponding to the curves in \figref{phasespace}. The response of the velocity field to the matter is different in GR and modified gravity. The difference is directly correlated to the difference between the two potentials $\Psi$ (which governs dynamics) and $\Phi$ (which governs the curvature). It is interesting to note that while $\Phi$ and $\Psi$ vary along the radial direction (see \figref{staticsoln}), their ratio is a constant and directly correlated with deviation of the DVDR curve from its GR counterpart in phasespace. }
\label{deviation}
\end{figure}

There are various parameters that have been used in the literature to characterize the departure of modified gravity models from GR (see for example \citealt{silvestri_practical_2013,lombriser_parametrisation_2016}) and it is useful to correlate the departure of the DVDR relation from GR with such parameters. We define the parameter (in real space) 
\beq 
\eta = \frac{\Psi}{\Phi}. 
\eeq
This is distinguished from $\eta_k = \frac{\Phi_k}{\Psi_k}$ defined by \cite{pogosian_pattern_2008} which is ratio of the Fourier components of $\Phi$ and $\Psi$. $\eta_k$ varies smoothly from $0.5$ to 1 while traversing the limit from modified gravity to GR. We found that the real space $\eta$ was constant along the radial direction for the entire range of the perturbation rising sharply to $\infty$ after the compensated region where $\Psi \sim \Phi \sim 0$, since it is ill-defined on that domain. In comparing the deviations, we consider only the domain of the perturbation where $\eta <1$. In \figref{deviation}, we plot the deviation of the curves in phasespace defined as follows. At a given radial point $x$, the pair $\{\Delta, \delta_{v,fR}\}$ is well-defined from the numerical density and velocity profiles evolved in the $f(R)$ theory and plotted in \figref{denvel}. The pair $\{\Delta, \delta_{v,GR}\}$ corresponds to the expected value of $\delta_v$ if the evolution assumed standard GR. The latter was computed by evolving the same initial profile in standard GR and interpolating the resulting $\Delta-\delta_v$ curve; it can also be computed using the fit given in N13.  The deviation between the two curves is given by 
\beq 
{\rm deviation}(x) = |\delta_{v,GR}(\Delta(x)) - \delta_{v,fR}(\Delta(x))|.
\eeq
The red and blue lines correspond to profiles I and II respectively. It is clear that there is a direct correspondence between $\eta$ and the deviation in phasespace. Note that $\eta$ and the deviation both remain approximately constant throughout the radius of the perturbation, although the density, velocity and potentials are radially varying. As expected, $\eta$ closer to unity implies as smaller deviation. 
 
\subsection{Fitting form for the DVDR in the strong field regime.} 
The non-linear density velocity divergence relation has been extensively investigated in the past using perturbation theory both in the Eulerian \citep{ bernardeau_quasi-gaussian_1992,chodorowski_large-scale_1997,chodorowski_weakly_1997} and Lagrangian frames \citep{susperregi_cosmic_1997} as well as  numerical simulations \citep{bernardeau_new_1996, chodorowski_recovery_1998, zaroubi_wiener_1999,kudlicki_reconstructing_2000,ciecielg_gaussianity_2003,kitaura_estimating_2012}. The formula given by \cite{bernardeau_quasi-gaussian_1992} (B92 hereafter) in standard GR reads 
\beq 
\Theta = \frac{3\Omega_m^{0.6}}{2} \left\{1-(1+\delta)^{2/3}\right \}
\eeq 
and is valid in the regime $-1\leq \delta \leq  2$. \cite{bilicki_velocity-density_2008}, BC08 hereafter, gave a fitting form valid for a larger range of $\delta$ but is based on the exact analytic solution of the spherical collapse and hence was derived for pure matter cosmologies. N13 gave a formula based on the B92 and BC08, but with the growth index modified to account for a general dark energy component with a constant equation of state $w$. \citet{mandal_one-point_2020} checked that this formula also holds for early dark energy scenarios where $w$ may vary with time.

In this paper we find a fitting form for DVDR in the strong field regime of $f(R)$ where the dynamics is scale independent. We evaluated the $\Delta-\delta_v$ pairs for final density and velocity profiles in the range  from  $0.5 \leq a \leq1$ using a function of the form $\delta_v = A(\Omega_m)  [1-(1+\Delta)^{B(\Omega_m})]$ where $A$ and $B$ were coefficients which depend on the instantaneous $\Omega_m$. $B$ was found to be independent of $\Omega_m$, whereas $A$ varies as a power law. The form 
\beq
\delta_v = 0.64 \times \Omega_m^{0.54} [1-(1+\Delta)^{0.61}]
\label{DVDRfit}
\eeq
gave a good fit over the range $0\leq \Delta \leq 2$ and $0.5 \leq a \leq1$. 
In terms of $\delta$ and $\Theta$ this relation reads 
\beq
\Theta = 1.92 \times \Omega_m^{0.54} [1-(1+\delta)^{0.61}]. 
\eeq
The r.m.s. relative error between the fit and the data was about 17\% over the entire range of $\Delta$ and $a$. In this regime, we do not expect the non-linear DVDR to depend on the exact choice of $a_{switch}$.
We did not attempt to fit the void region since the top-hats considered here are overdense giving a positive $\Delta$. 

\begin{figure} 
\includegraphics[width=18cm]{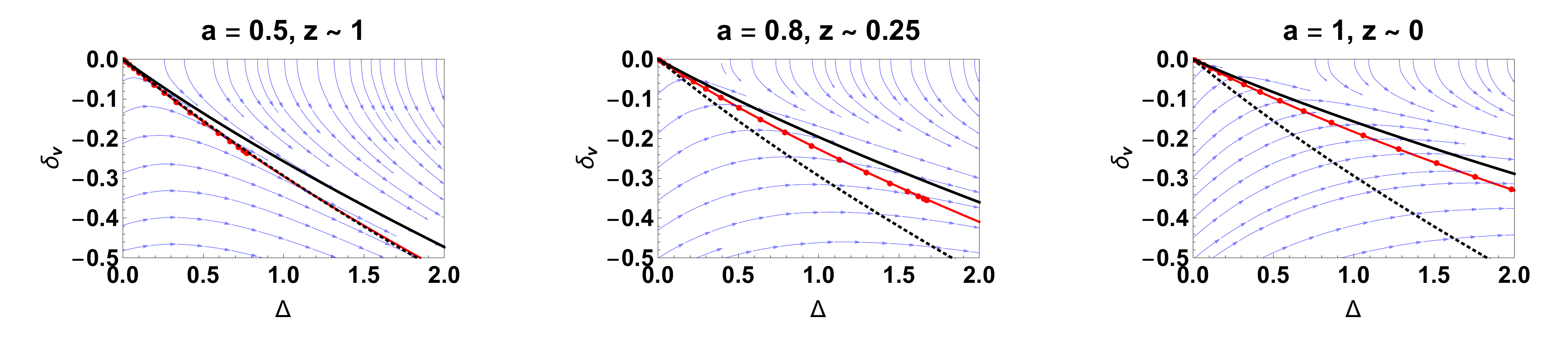}
\caption{The DVDR in the strong field limit ($Q>>1$). The red line is the fitting form given in \eqnref{DVDRfit} and the red dots show the $\Delta-\delta_v$ pairs computed from the radial profile at the selected epoch. The solid black line shows the DVDR for the reference $\Lambda$CDM model considered here and the black dotted line shows the DVDR for the flat EdS model with $\Omega_m =1$. The dynamical system of equations governing $\Delta$ and $\delta_v$ is non-autonomous in the case of $\Lambda$CDM and $f(R)$ and hence the curves are time-dependent. On the other hand, the EdS curve is the same for all epochs.  }
\label{fit}
\end{figure}

\capfigref{fit} shows the fitting form for the DVDR relation for the $\Lambda$CDM case (black) and the $f(R)$ model (red) in the strong field regime as a function of redshift. The red points represent the numerically evolved $(\Delta, \delta_v)$ points corresponding to the selected epoch. The blue arrows show the streamlines of the flow given by \eqnrefs{MGA2} and \eqnrefbare{MGB2}. They show the instantaneous direction of the vector defined by $(\Delta', \delta_v')$. The system defined by  \eqnrefs{MGA2} and \eqnrefbare{MGB2} is a non-autonomous system since the coefficients of $\Delta$ and $\delta_v$ in the r.h.s. are time-dependent. Hence, the streamlines are not constant at all epochs and neither is the DVDR curve. This can be contrasted with the case of a pure $\Omega_m =1$ flat universe, where the dynamical system for $\Delta-\delta_v$ is an autonomous system and the DVDR relation is time-independent (see N13 for a detailed discussion). 

The DVDR relation based on spherical collapse is one-to-one. However, for realistic density and velocity fields, there is a scatter around a mean relation. Numerical simulations in the case of $\Lambda$CDM have shown that the mean relation is close to that given by the spherical collapse model \citep{bernardeau_non-linearity_1999,kudlicki_reconstructing_2000, bilicki_velocity-density_2008}. This scatter is usually attributed to the non-local nature of gravity. The scatter can be related to the shear component of the velocity field as was shown by \cite{chodorowski_large-scale_1997} using perturbation theory and using triaxial collapse by \cite{nadkarni-ghosh_phase_2016}. It remains to be checked whether the mean relation from the joint density-velocity PDF obtained from simulations agrees with the estimate based on spherical collapse given in this paper. 
\subsection{Density enhancement in non-linear evolution}
\label{sec:highNL}
\begin{figure}
\includegraphics[width=16cm]{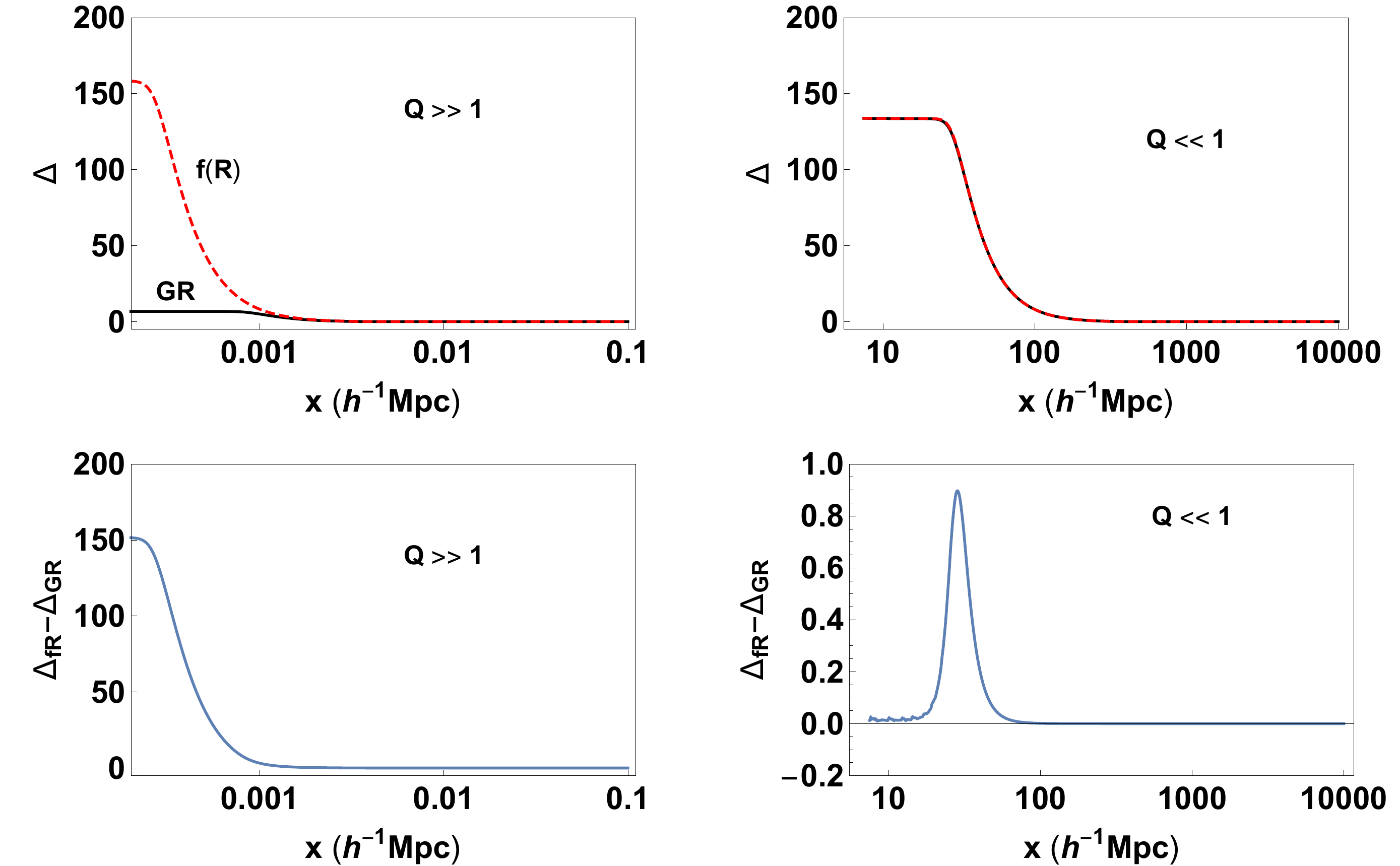}
\caption{Demonstration of non-linear growth in strong vs. weak regimes. In the upper panel, the red dashed denotes $f(R)$ and black solid is the GR value for $\Delta$ at $a=1$. In the strong regime $Q>>1$, the profile is amplified as compared to GR, whereas for the weak field regime, where $Q<<1$, the two profiles coincide by eye. The lower panel  profile develops a cusp at the edge of the top-hat. This is related to the solution for $\chi$ which is proportional to the density in the weak field limit. Since the force is proportional to the $\nabla \chi$, this manifests itself as a cusp at the edge of the top-hat.  }
\label{highNL}
\end{figure}

In this paper, we do not undertake a systematic study of the highly non-linear regime, in particular the process of collapse. However, we illustrate, qualitatively, the difference between highly non-linear evolution in the strong and weak limits. We consider two profiles: both with $\sigma = 0.0025$, but with different $x_{top}$. One case corresponds to $x_{top} = 0.002 h^{-1}$Mpc and the other with $x_{top} = 200 h^{-1}$Mpc. $Q>>1$ for the former and $Q<<1$ for the latter over the entire range of evolution from $a=0.1$ to $a=1$. In each case, the initial amplitude was chosen so that the final evolved density in the $f(R)$ model has $\delta \sim 100$. This implies that in the strong field limit, the initial amplitude of the perturbation is less than that for the weak field limit. The details are in appendix \ref{app:tophatsetup}. The resulting non-linear $\Delta$ is shown in the top-panel of \figref{highNL} and the difference between the $f(R)$ and GR profiles is shown in the bottom panel. For $Q>>1$, 
the amplification is large since the effective force is $4/3$ times stronger. For $Q<<1$, the evolved profiles for $f(R)$ and GR apparently coincide. However, the lower panel shows that the $f(R)$ model has a higher density at the edge of the top-hat. This can be explained in terms of the solution for $\chi$ as was discussed in \S \ref{sec:potentials}. $\chi$ is proportional to $\delta$ in this regime and the force which is proportional to the gradient is non-zero only at the edge of the top-hat.

The density enhancement at the edge has been reported earlier \citep{borisov_spherical_2012,kopp_spherical_2013, lombriser_parametrisation_2016} but in the presence of the chameleon screening. The appearance of a `spike' at the edge of the top-hat is akin to the presence of a  `thin-shell' in the chameleon mechanism, although in this paper we have not explicitly modelled the chameleon screening. Hence, it is worth understanding the relation between these two features. Following the seminal paper on `chameleon cosmology' \citep{khoury_chameleon_2004}, we consider a scalar field $\phi$, evolving according to 
\beq 
\nabla_r^2 \phi = V_{eff}(\phi)_{,\phi}, 
\eeq 
where $V_{eff} =V(\phi) + \rho e^{\sqrt{8 \pi G} \beta \phi} $ and $\beta$ is a coupling constant of order unity. This equation is solved for an overdense sphere of density $\rho$ embedded in a background of density $\rho_\infty< \rho$, with boundary conditions $\phi (r \rightarrow \infty) = \phi_\infty$, where $\phi_\infty$ is the value of the field that minimises the effective potential at $\infty$ and $d \phi/dr ( r \rightarrow 0) = 0$. The `chameleon solution' refers to a special solution of this equation where the effective potential is at its minimum {\it almost} everywhere inside the sphere, except for a thin shell at the surface of the sphere. This solution corresponds to $\nabla_r^2 \phi \approx V_{eff, \phi} \approx 0$ and the `extra fifth force', which is given by $\nabla \phi$ is zero inside the sphere except near the surface. When $\rho$ is large, as is the case inside the sphere,  the second derivative of the effective potential at the minimum is larger i.e., the potential is steeper. The 
mass of the scalar field for this solution is large and correspondingly, the Compton wavelength is small inside the sphere. Outside the sphere, the density is low, the effective potential is shallower, the Compton wavelength is large and the fifth force is not screened.

In the $f(R)$ model, the scalar field corresponds to $\chi = \delta f_R$. The $Q << 1$ solution corresponds to $\nabla_x^2 \chi \approx 0$, which gives $\chi =  {\bar x}_c^2 H^2 \Omega_m a^2 \delta$. Since the `fifth force' depends on $\nabla_x \chi$, it is zero everywhere except at the edge of the top-hat, where the derivative of the density is non-zero. However, we do not refer to this as `chameleon screening' since the Compton wavelength is constant both inside and outside the top-hat and is independent of $\delta$. This can also be seen by defining an `effective potential' $V_{eff}(\chi) = \frac{\chi^2}{2{\bar x}_C^2} + H^2 \Omega_m a^2 \delta \chi$, recasting the equation for $\chi$ as $\nabla_x^2 \chi = V_{eff}(\chi)_{,\chi}$ and noting that the second derivative, $V_{eff, \chi \chi}$, which is inversely related to the Compton wavelength, is independent of $\delta$. In this paper, we have used the first order of the Taylor expansion $\delta f_R \sim f_{RR} \delta R + \mathcal{O}(R)$ in obtaining a linear equation for $\chi$. Ideally, as has been argued by Hu-Sawicki, this linearization is not valid when $\delta R$ is large (high curvature limit). Alternately, it is clear that when the Compton wavelength is small i.e., $f_{RR}$ is small, higher order corrections to this expansion may become important making the equation for $\chi$ non-linear. Higher order terms can be modelled by allowing the Compton wavelength to depend on $\chi$. Thus, the chameleon solution is essentially a solution of the non-linear equation for $\chi$. 

Physically, the density spike at the edge is related to the enhanced force on the shells near the edge. For a constant density top-hat, all shells up to the edge experience the same acceleration and the top-hat maintains its shape until the innermost shell collapses first. On the other hand, in $f(R)$ gravity, even though at the initial time, all shells start with the same acceleration, as evolution proceeds, the shells near the edge accelerate faster than they would in GR. The inner shells accelerate at the same rate as GR since the fifth force is screened inside. Mass starts to accumulate at the edge giving rise to an enhanced density, until, eventually, the faster moving shells cross the inner slower shells giving rise to a caustic at the edge. This was observed in \cite{borisov_spherical_2012}. It is important to note that this density enhancement is pronounced for a top-hat due to the presence of a steep gradient. The ratio of the additional force to the GR force is 
 \beq 
 \frac{\nabla_x \chi}{\nabla_x \Phi_{GR}} \sim {\bar x}_C \frac{d \ln \delta}{dx}. 
 \eeq
 Enhancement will be prominent only when the density changes over a scale much smaller than the Compton wavelength of the scalar field. For a smooth radial density, enhancement will be minimal in the weak field regime. Thus, the edge effect is observed when the scale of the density perturbation is larger than the Compton wavelength, but the scale of the density gradients are larger than the Compton wavelength. 
 In the strong field regime, when $Q>>1$, the the Compton wavelength is large compared to the scale of the perturbation. There are no edge effects since the `fifth force' and the associated density enhancement is uniform; the final density is higher as compared to GR. When $Q \sim 1$, there is a slight edge effect (see middle panel of $\nabla \chi$ in \figref{staticsoln}), but is suppressed compared to the $Q<<1$ case because the Compton wavelength is larger. Understanding this behaviour for the full non-linear equation is left for future work. 
 
\section{Summary and discussion}
\label{sec:conclusion}
In this work, we have investigated the non-linear evolution of cosmological perturbations in $f(R)$ models using the compensated spherical top-hat as a proxy for the non-linear regime. Our algorithm is an iterative hybrid Lagrangian-Eulerian scheme; the continuity equation and Euler equation are solved in Lagrangian coordinates and the equations for the metric potentials are purely spatial and are solved analytically in Eulerian space. Specifically, we chose the  \cite{hu_models_2007} model for simplicity, with majority of the results obtained by setting $f_{R0} = -10^{-6}$ and $n=1$. The evolution equations depend upon the ratio of $ Q = {\bar x_C}/{x_{top}}$, where ${\bar x_C}$ is the reduced comoving Compton wavelength of the model given by \eqnref{Comptondef2}. The Compton wavelength is assumed to depend only on the background cosmology; thus we do not model the chameleon screening mechanism (\citealt{khoury_chameleon_2004}). However, we demonstrate the role played by the Compton wavelength by changing the scale of the top-hat ($x_{top}$).  We consider three values of $x_{top} =0.002, 2$ and 200 $h^{-1}{\rm Mpc}$, corresponding to $Q = 1217, 1.21$ and 0.012 at $a=1$. 
There are several new features of this work. 
\begin{itemize} 
\item {\bf Analysis of oscillations in the background evolution}\\
It is well known that the $f(R)$ model exhibits high frequency oscillations at high redshifts, both in the background and perturbation evolution making the system numerically stiff. To overcome this, it is favourable to assume GR at early epochs and `switch' to $f(R)$ evolution at a late epoch denoted by $a_{switch}$.  By performing an eigenvalue analysis of the linearized background equations, we were able to assess the nature of oscillations and make an informed choice of $a_{switch}$. Such an eigenanalysis has been performed for the Boltzmann hierarchy in order to gain insights for optimizing solvers (\citealt{nadkarni-ghosh_einstein-boltzmann_2017}) 
and a similar analysis can also be performed for the linear perturbation system of $f(R)$ or other models of modified gravity. In particular, it could help in putting the Quasi-Static-Approximation on firmer grounds (\citealt{Hojjati_2012,sawicki_limits_2015,pace_comparison_2021}). 

\item {\bf Analytic solutions for the metric potentials and density-enhancement}\\
Growth of perturbations in standard GR is governed by the joint continuity, Euler and Poisson equations. In $f(R)$ models, there is an extra equation corresponding to the `extra' degree of freedom, usually encapsulated by the variable $f_R$. In the limit $|f_R|<<1$ and $|f/R|<<1$, the perturbations in $f_R$ can be expressed as the difference of the two potentials $\Phi - \Psi$, where $\Phi$ and $\Psi$ are the metric perturbations corresponding to the `curvature' potential and `Newtonian' potential respectively. We derive an analytic solution for $\chi = \Phi - \Psi$ in real space in spherical symmetry valid for all values of $Q$. In the strong field limit, when the scale of the perturbation is much less than its Compton wavelength ($Q>>1$), $\chi$ obeys the Poisson equation and $\Psi$ is 4/3 times its value in standard GR. In the weak field limit, when the scale of the perturbation is much greater than its Compton wavelength ($Q<<1$), $\chi \propto \delta$. This means that the force, which is proportional to the gradient of $\chi$ is non-zero only at the edge of the top-hat. This results in a density enhancement at the top-hat edge, a phenomenon that has been previously reported in the literature \citep{borisov_spherical_2012, kopp_spherical_2013} in the context of `chameleon screening'.  Here we demonstrate the same effect in the absence of a chameleon mechanism. If the Compton wavelength is allowed to change within the perturbation, then the resulting equation for $\chi$ is non-linear, possibly enhancing this effect. 
A more quantitative exploration is left for future investigations. Nevertheless, the analytic solutions presented here can also serve as a `test case' to check other numerical schemes, similar in spirit to the use of spherical collapse as a `control case' for numerical simulations  (see \citealt{joyce_evolution_2012}).

\item {\bf Phasespace analysis of evolution of perturbations and the density-velocity divergence DVDR relation in the strong field limit}\\
We evolve the compensated top-hats and examine the evolution of the perturbations in the 2-D density-velocity divergence phase space. For the radially varying profiles, it is easier to compute the $\Delta-\delta_v$ paris instead of the $\delta-\Theta$ pairs, where $\Delta$ is the spherically averaged density and $\delta_v$ is the fractional Hubble velocity of the shell. In the strong field limit, where the evolution of perturbations is scale-invariant, it is possible to define a unique DVDR relation and we give a fitting form for the same. In the intermediate field limit, $Q \sim 1$, the evolution of perturbations is profile dependent and an invariant relation between $\delta$ and $\Theta$ does not exist. In the weak field limit, the dynamics is approximately that of GR.
\end{itemize}

There are several caveats in the present work. The spherical collapse is a simplistic model: the system does not take into account any effect of the environment or rotations and ignores mode-mode coupling. The first step forward is to consider triaxial collapse. Triaxial collapse, which is an important ingredient in mock catalog generators like PINOCCHIO \citep{2016monaco,Rizzo_2017,Moretti_2020,song_2021} has not been studied in great detail in the context of modified gravity, albeit, with the exception of some recent efforts  (\citealt{Burrage_2015,Burrage_2018,Ruan_2020}). To capture the effects of mode coupling in the analytic framework, it is necessary to consider perturbation theory, either in the Eulerian or Lagrangian frame and related schemes (see e.g., \citealt{Aviles_2017, Aviles_2019,Valogiannis_2020,Aviles_2021, li_approximation_2018}). The hybrid Eulerian-Lagrangian scheme outlined here does not rely on spherical geometry. \capeqnrefs{conteq}-\eqnrefbare{chi} are general and the hybrid scheme can also be incorporated with a multi-step Lagrangian perturbation theory \citep{nadkarni-ghosh_extending_2011,nadkarni-ghosh_modelling_2013}, which guarantees convergence in $\Lambda$CDM models up until shell crossing. However, convergence and shell-crossing related issues in modified gravity scenarios remain to be investigated (see for example \citealt{Rampf_2017, Rampf_2020} for a discussion of these matters in standard GR). A code, called SELCIE,  based on Finite Element Methods to solve the chameleon equations of motion in arbitrary mass distributions has recently been developed by \cite{Briddon:2021etm}. However, it is currently not set up to perform temporal evolution of the matter distribution in the presence of chameleon screening. It may be possible to couple iterative methods used in this paper with such tools. In addition, for the $f(R)$ model considered here, the relevant scales for which the system is in the strong field limit, are not cosmological scales, but correspond to smaller bodies governed by additional astrophysical processes which have also been ignored in this treatment. 

The $f(R)$ model, in the metric formalism can be considered as a special case of Brans-Dickie theories, which in turn belongs to the more general class of Horndeski theories \citep{defelice_fR_2010, kobayashi_2019}. More recently, the Effective Field Theory of dark energy has become a popular formalism which is aimed at capturing a wide class of modified theory models (see review by \citealt{perenon_2020}). 
Other kinds of $f(R)$ modifications have also been introduced in the context of understanding galaxy rotation curves (for e.g., \citealt*{Dey_2013,
Dey_2015}) or in the context of unified models of dark energy and inflation (for e.g.,  \citealt{Nojiri_2007,Cognola_2008,Nojiri_2011}). Spherically symmetric solutions play an important role in understanding the collapse processes in such systems \citep{astashenok_2019,chowdhury_collapse_2020,Jaryal_2021}. Furthermore, many future tests of modified gravity will involve measurements on a very wide range of astrophysical and cosmological scales depending upon the magnitude of modification to the `curvature' and `potential' fields \citep{Baker_2015}. Many of these systems are often modelled by spherical symmetry. We hope that some of the techniques presented in this work may be extended to gain insights into these generalized models of modified gravity in such systems. 


\section{Acknowledgements}
S.N. would like to acknowledge the Department of Science and Technology, Govt. of India, for the grant no. WOS-A/PM-21/2018. S.C. would like to acknowledge the Council of Scientific \& Industrial Research (CSIR) for the grant no. 09/092(0930)/2015-EMR-I. Both authors would like to thank Tapobrata Sarkar for useful discussions through the course of this project and for detailed comments on the manuscript. S. N. would like to thank Alessandra Silvestri for pointing out some useful references. 

\section{Data Availability}
Data generated by the numerical runs available on request from the corresponding author. 
\newpage
\bibliographystyle{mn2e.bst}
\bibliography{modgrav,book}

\begin{thebibliography}{124}
\expandafter\ifx\csname natexlab\endcsname\relax\def\natexlab#1{#1}\fi

\bibitem[{{Alam} {et~al}\mbox{.}(2017){Alam}, {Ata}, {Bailey}, {Beutler},
  {Bizyaev}, {Blazek}, {Bolton}, {Brownstein}, {Burden}, {Chuang}, {Comparat},
  {Cuesta}, {Dawson}, {Eisenstein}, {Escoffier}, {Gil-Mar{\'\i}n}, {Grieb},
  {Hand}, {Ho}, {Kinemuchi}, {Kirkby}, {Kitaura}, {Malanushenko},
  {Malanushenko}, {Maraston}, {McBride}, {Nichol}, {Olmstead}, {Oravetz},
  {Padmanabhan}, {Palanque-Delabrouille}, {Pan}, {Pellejero-Ibanez},
  {Percival}, {Petitjean}, {Prada}, {Price-Whelan}, {Reid},
  {Rodr{\'\i}guez-Torres}, {Roe}, {Ross}, {Ross}, {Rossi},
  {Rubi{\~n}o-Mart{\'\i}n}, {Saito}, {Salazar-Albornoz}, {Samushia},
  {S{\'a}nchez}, {Satpathy}, {Schlegel}, {Schneider}, {Sc{\'o}ccola}, {Seo},
  {Sheldon}, {Simmons}, {Slosar}, {Strauss}, {Swanson}, {Thomas}, {Tinker},
  {Tojeiro}, {Maga{\~n}a}, {Vazquez}, {Verde}, {Wake}, {Wang}, {Weinberg},
  {White}, {Wood-Vasey}, {Y{\`e}che}, {Zehavi}, {Zhai}, \&
  {Zhao}}]{alam_SDSS_2017}
{Alam} S. {et~al.}, 2017, Monthly Notices of the Royal Astronomical Society,
  470, 2617

\bibitem[{Alam {et~al}\mbox{.}(2020)Alam, Aviles, Bean, Cai, Cautun,
  Cervantes-Cota, Cuesta-Lazaro, Chandrachani~Devi, Eggemeier, Fromenteau,
  Gonzalez-Morales, Halenka, He, Hellwing, Hernandez-Aguayo, Ishak, Koyama, Li,
  de~la Macorra, Menesses~Rizo, Miller, Mueller, Niz, Ntelis, Rodriguez~Otero,
  Sabiu, Slepian, Stark, Valenzuela, Valogiannis, Vargas-Magana, Winther,
  Zarrouk, Zhao, \& Zheng}]{alam_testing_2020}
Alam S. {et~al.}, 2020, arXiv e-prints, 2011, arXiv:2011.05771

\bibitem[{Amendola {et~al}\mbox{.}(2018)Amendola, Appleby, Avgoustidis, Bacon,
  Baker, Baldi, Bartolo, Blanchard, Bonvin, Borgani, Branchini, Burrage,
  Camera, Carbone, Casarini, Cropper, de~Rham, Dietrich, Di~Porto, Durrer,
  Ealet, Ferreira, Finelli, GarcÃ­a-Bellido, Giannantonio, Guzzo, Heavens,
  Heisenberg, Heymans, Hoekstra, Hollenstein, Holmes, Hwang, Jahnke, Kitching,
  Koivisto, Kunz, La~Vacca, Linder, March, Marra, Martins, Majerotto, Markovic,
  Marsh, Marulli, Massey, Mellier, Montanari, Mota, Nunes, Percival, Pettorino,
  Porciani, Quercellini, Read, Rinaldi, Sapone, Sawicki, Scaramella, Skordis,
  Simpson, Taylor, Thomas, Trotta, Verde, Vernizzi, Vollmer, Wang, Weller, \&
  Zlosnik}]{amendola_2018}
Amendola L. {et~al.}, 2018, Living Reviews in Relativity, 21, 2

\bibitem[{Appleby \& Battye(2007)}]{appleby_consistent_2007}
Appleby S., Battye R., 2007, Physics Letters B, 654, 7

\bibitem[{Astashenok {et~al}\mbox{.}(2019)Astashenok, Mosani, Odintsov, \&
  Samanta}]{astashenok_2019}
Astashenok A.~V., Mosani K., Odintsov S.~D., Samanta G.~C., 2019, International
  Journal of Geometric Methods in Modern Physics, 16, 1950035

\bibitem[{Aviles \& Cervantes-Cota(2017)}]{Aviles_2017}
Aviles A., Cervantes-Cota J.~L., 2017, Physical Review D, 96

\bibitem[{Aviles {et~al}\mbox{.}(2019)Aviles, Cervantes-Cota, \&
  Mota}]{Aviles_2019}
Aviles A., Cervantes-Cota J.~L., Mota D.~F., 2019, Astronomy \& Astrophysics,
  622, A62

\bibitem[{Aviles {et~al}\mbox{.}(2021)Aviles, Valogiannis, Rodriguez-Meza,
  Cervantes-Cota, Li, \& Bean}]{Aviles_2021}
Aviles A., Valogiannis G., Rodriguez-Meza M.~A., Cervantes-Cota J.~L., Li B.,
  Bean R., 2021, Journal of Cosmology and Astroparticle Physics, 2021, 039

\bibitem[{Baker {et~al}\mbox{.}(2021)Baker, Barreira, Desmond, Ferreira, Jain,
  Koyama, Li, Lombriser, Nicola, Sakstein, \& Schmidt}]{baker_novel_2021}
Baker T. {et~al.}, 2021, Reviews of Modern Physics, 93, 015003

\bibitem[{Baker {et~al}\mbox{.}(2015)Baker, Psaltis, \& Skordis}]{Baker_2015}
Baker T., Psaltis D., Skordis C., 2015, The Astrophysical Journal, 802, 63

\bibitem[{Baojiu(2018)}]{li_approximation_2018}
Baojiu L., 2018, International Journal of Modern Physics, 27

\bibitem[{Bardeen {et~al}\mbox{.}(1986)Bardeen, Bond, Kaiser, \&
  Szalay}]{bardeen_statistics_1986}
Bardeen J.~M., Bond J.~R., Kaiser N., Szalay A.~S., 1986, The Astrophysical
  Journal, 304, 15

\bibitem[{Barreira {et~al}\mbox{.}(2013)Barreira, Li, Baugh, \&
  Pascoli}]{barreira_spherical_2013}
Barreira A., Li B., Baugh C.~M., Pascoli S., 2013, Journal of Cosmology and
  Astro-Particle Physics, 11, 056

\bibitem[{Bean {et~al}\mbox{.}(2007)Bean, Bernat, Pogosian, Silvestri, \&
  Trodden}]{bean_dynamics_2007}
Bean R., Bernat D., Pogosian L., Silvestri A., Trodden M., 2007, Physical
  Review D, 75, 064020

\bibitem[{Bernardeau(1992)}]{bernardeau_quasi-gaussian_1992}
Bernardeau F., 1992, The Astrophysical Journal, 390, L61

\bibitem[{Bernardeau {et~al}\mbox{.}(1999)Bernardeau, Chodorowski, Lokas,
  Stompor, \& Kudlicki}]{bernardeau_non-linearity_1999}
Bernardeau F., Chodorowski M.~J., Lokas E.~L., Stompor R., Kudlicki A., 1999,
  Monthly Notices of the Royal Astronomical Society, 309, 543

\bibitem[{Bernardeau \& van~de Weygaert(1996)}]{bernardeau_new_1996}
Bernardeau F., van~de Weygaert R., 1996, Monthly Notices of the Royal
  Astronomical Society, 279, 693

\bibitem[{Bertschinger \& Dekel(1989)}]{bertschinger_recovering_1989}
Bertschinger E., Dekel A., 1989, The Astrophysical Journal, 336, L5

\bibitem[{Bertschinger \& Zukin(2008)}]{bertschinger_distinguishing_2008}
Bertschinger E., Zukin P., 2008, Physical Review D, 78, 024015

\bibitem[{Bilicki \& Chodorowski(2008)}]{bilicki_velocity-density_2008}
Bilicki M., Chodorowski M.~J., 2008, Monthly Notices of the Royal Astronomical
  Society, 391, 1796

\bibitem[{Borisov {et~al}\mbox{.}(2012)Borisov, Jain, \&
  Zhang}]{borisov_spherical_2012}
Borisov A., Jain B., Zhang P., 2012, Physical Review D, 85, 63518

\bibitem[{Brenier {et~al}\mbox{.}(2003)Brenier, Frisch, H\'{e}non, Loeper,
  Matarrese, Mohayaee, \& Sobolevsk\~{ii}}]{brenier_reconstruction_2003}
Brenier Y., Frisch U., H\'{e}non M., Loeper G., Matarrese S., Mohayaee R.,
  Sobolevsk\~{ii} A., 2003, Monthly Notices of the Royal Astronomical Society,
  346, 501

\bibitem[{Briddon {et~al}\mbox{.}(2021)Briddon, Burrage, Moss, \&
  Tamosiunas}]{Briddon:2021etm}
Briddon C., Burrage C., Moss A., Tamosiunas A., 2021, Journal of Cosmology and
  Astroparticle Physics, 2021, 043

\bibitem[{Buchert(1992)}]{buchert_92}
Buchert T., 1992, Monthly Notices of the Royal Astronomical Society, 254, 729

\bibitem[{Burrage {et~al}\mbox{.}(2018)Burrage, Copeland, Moss, \&
  Stevenson}]{Burrage_2018}
Burrage C., Copeland E.~J., Moss A., Stevenson J.~A., 2018, Journal of
  Cosmology and Astroparticle Physics, 2018, 056?056

\bibitem[{Burrage {et~al}\mbox{.}(2015)Burrage, Copeland, \&
  Stevenson}]{Burrage_2015}
Burrage C., Copeland E.~J., Stevenson J.~A., 2015, Physical Review D, 91

\bibitem[{Cataneo {et~al}\mbox{.}(2021)Cataneo, Uhlemann, Arnold, Gough, Li, \&
  Heymans}]{cataneo2021matter}
Cataneo M., Uhlemann C., Arnold C., Gough A., Li B., Heymans C., 2021, The
  matter density pdf for modified gravity and dark energy with large deviations
  theory

\bibitem[{Ceron-Hurtado {et~al}\mbox{.}(2016)Ceron-Hurtado, He, \&
  Li}]{ceron-hurtado_can_2016}
Ceron-Hurtado J.~J., He J.-h., Li B., 2016, Physical Review D, 94, 064052

\bibitem[{Chakrabarti \& Banerjee(2016)}]{chakrabarti_spherically_2016}
Chakrabarti S., Banerjee N., 2016, General Relativity and Gravitation, 48, 57

\bibitem[{Chakraborty {et~al}\mbox{.}(2021)Chakraborty, MacDevette, \&
  Dunsby}]{chakraborty_model_2021}
Chakraborty S., MacDevette K., Dunsby P., 2021, arXiv e-prints, 2103,
  arXiv:2103.02274

\bibitem[{Chiba(2003)}]{chiba_1r_2003}
Chiba T., 2003, Physics Letters B, 575, 1

\bibitem[{Chodorowski(1997)}]{chodorowski_large-scale_1997}
Chodorowski M.~J., 1997, Monthly Notices of the Royal Astronomical Society,
  292, 695

\bibitem[{Chodorowski \& Lokas(1997)}]{chodorowski_weakly_1997}
Chodorowski M.~J., Lokas E.~L., 1997, Monthly Notices of the Royal Astronomical
  Society, 287, 591

\bibitem[{Chodorowski {et~al}\mbox{.}(1998)Chodorowski, Lokas, Pollo, \&
  Nusser}]{chodorowski_recovery_1998}
Chodorowski M.~J., Lokas E.~L., Pollo A., Nusser A., 1998, Monthly Notices of
  the Royal Astronomical Society, 300, 1027

\bibitem[{{Chowdhury} {et~al}\mbox{.}(2020){Chowdhury}, {Pal}, {Pal}, \&
  {Sarkar}}]{chowdhury_collapse_2020}
{Chowdhury} S., {Pal} K., {Pal} K., {Sarkar} T., 2020, European Physical
  Journal C, 80, 902

\bibitem[{Ciecielg {et~al}\mbox{.}(2003)Ciecielg, Chodorowski, Kiraga, Strauss,
  Kudlicki, \& Bouchet}]{ciecielg_gaussianity_2003}
Ciecielg P., Chodorowski M.~J., Kiraga M., Strauss M.~A., Kudlicki A., Bouchet
  F.~R., 2003, Monthly Notices of the Royal Astronomical Society, 339, 641

\bibitem[{Clifton {et~al}\mbox{.}(2012)Clifton, Ferreira, Padilla, \&
  Skordis}]{clifton_modified_2012}
Clifton T., Ferreira P.~G., Padilla A., Skordis C., 2012, Physics Reports, 513,
  1

\bibitem[{Cognola {et~al}\mbox{.}(2008)Cognola, Elizalde, Nojiri, Odintsov,
  Sebastiani, \& Zerbini}]{Cognola_2008}
Cognola G., Elizalde E., Nojiri S., Odintsov S.~D., Sebastiani L., Zerbini S.,
  2008, Physical Review D, 77

\bibitem[{Colombi {et~al}\mbox{.}(2007)Colombi, Chodorowski, \&
  Teyssier}]{colombi_cosmic_2007}
Colombi S., Chodorowski M.~J., Teyssier R., 2007, Monthly Notices of the Royal
  Astronomical Society, 375, 348

\bibitem[{Dai {et~al}\mbox{.}(2008)Dai, Maor, \&
  Starkman}]{dai_consequences_2008}
Dai D.-C., Maor I., Starkman G., 2008, Physical Review D, 77, 64016

\bibitem[{{De Felice} \& {Tsujikawa}(2010)}]{defelice_fR_2010}
{De Felice} A., {Tsujikawa} S., 2010, Living Reviews in Relativity, 13, 3

\bibitem[{Dey {et~al}\mbox{.}(2013)Dey, Bhattacharya, \& Sarkar}]{Dey_2013}
Dey D., Bhattacharya K., Sarkar T., 2013, Physical Review D, 87

\bibitem[{Dey {et~al}\mbox{.}(2015)Dey, Bhattacharya, \& Sarkar}]{Dey_2015}
Dey D., Bhattacharya K., Sarkar T., 2015, General Relativity and Gravitation,
  47

\bibitem[{Di~Valentino {et~al}\mbox{.}(2021)Di~Valentino, Mena, Pan, Visinelli,
  Yang, Melchiorri, Mota, Riess, \& Silk}]{Di_Valentino_2021}
Di~Valentino E. {et~al.}, 2021, Classical and Quantum Gravity, 38, 153001

\bibitem[{Dodelson(2003)}]{dodelson}
Dodelson S., 2003, Modern Cosmology. Academic Press

\bibitem[{Dvali {et~al}\mbox{.}(2000)Dvali, Gabadadze, \&
  Porrati}]{dvali_4d_2000}
Dvali G., Gabadadze G., Porrati M., 2000, Physics Letters B, 485, 208

\bibitem[{Elizalde {et~al}\mbox{.}(2012)Elizalde, Odintsov, Sebastiani, \&
  Zerbini}]{elizalde_oscillations_2012}
Elizalde E., Odintsov S.~D., Sebastiani L., Zerbini S., 2012, European Physical
  Journal C, 72, 1843

\bibitem[{Frusciante \& Perenon(2020)}]{perenon_2020}
Frusciante N., Perenon L., 2020, Physics Reports, 857, 1?63

\bibitem[{Gramann(1993{\natexlab{a}})}]{gramann_improved_1993}
Gramann M., 1993{\natexlab{a}}, The Astrophysical Journal, 405, 449

\bibitem[{Gramann(1993{\natexlab{b}})}]{gramann_second-order_1993}
Gramann M., 1993{\natexlab{b}}, The Astrophysical Journal Letters, 405, L47

\bibitem[{Hahn {et~al}\mbox{.}(2015)Hahn, Angulo, \&
  Abel}]{hahn_properties_2015}
Hahn O., Angulo R.~E., Abel T., 2015, Monthly Notices of the Royal Astronomical
  Society, 454, 3920

\bibitem[{Herrera {et~al}\mbox{.}(2017)Herrera, Waga, \&
  Jor\'{a}s}]{Herrera_2017}
Herrera D., Waga I., Jor\'{a}s S., 2017, Physical Review D, 95

\bibitem[{Hojjati {et~al}\mbox{.}(2012)Hojjati, Pogosian, Silvestri, \&
  Talbot}]{Hojjati_2012}
Hojjati A., Pogosian L., Silvestri A., Talbot S., 2012, Physical Review D, 86

\bibitem[{Hu \& Sawicki(2007)}]{hu_models_2007}
Hu W., Sawicki I., 2007, Physical Review D, 76, 104043

\bibitem[{Jaryal \& Chatterjee(2021)}]{Jaryal_2021}
Jaryal S.~C., Chatterjee A., 2021, The European Physical Journal C, 81

\bibitem[{Joyce {et~al}\mbox{.}(2015)Joyce, Jain, Khoury, \&
  Trodden}]{joyce_beyond_2015}
Joyce A., Jain B., Khoury J., Trodden M., 2015, Physics Reports, 568, 1

\bibitem[{Joyce {et~al}\mbox{.}(2016)Joyce, Lombriser, \&
  Schmidt}]{joyce_dark_2016}
Joyce A., Lombriser L., Schmidt F., 2016, Annual Review of Nuclear and Particle
  Science, 66, 95

\bibitem[{Joyce \& Labini(2012)}]{joyce_evolution_2012}
Joyce M., Labini F.~S., 2012, Monthly Notices of the Royal Astronomical
  Society, 429, 1088

\bibitem[{Khoury \& Weltman(2004)}]{khoury_chameleon_2004}
Khoury J., Weltman A., 2004, Physical Review D, 69, 044026

\bibitem[{Kitaura {et~al}\mbox{.}(2012)Kitaura, Angulo, Hoffman, \&
  Gottl\"{o}ber}]{kitaura_estimating_2012}
Kitaura F.-S., Angulo R.~E., Hoffman Y., Gottl\"{o}ber S., 2012, Monthly
  Notices of the Royal Astronomical Society, 425, 2422

\bibitem[{Kobayashi(2019)}]{kobayashi_2019}
Kobayashi T., 2019, Reports on Progress in Physics, 82, 086901

\bibitem[{Kopp {et~al}\mbox{.}(2013)Kopp, Appleby, Achitouv, \&
  Weller}]{kopp_spherical_2013}
Kopp M., Appleby S.~A., Achitouv I., Weller J., 2013, Physical Review D, 88,
  084015

\bibitem[{Kudlicki {et~al}\mbox{.}(2000)Kudlicki, Chodorowski, Plewa, \&
  R\'{o}\.{z}yczka}]{kudlicki_reconstructing_2000}
Kudlicki A., Chodorowski M., Plewa T., R\'{o}\.{z}yczka M., 2000, Monthly
  Notices of the Royal Astronomical Society, 316, 464

\bibitem[{Li \& Efstathiou(2012)}]{li_extended_2012}
Li B., Efstathiou G., 2012, Monthly Notices of the Royal Astronomical Society,
  421, 1431

\bibitem[{Li \& Lam(2012)}]{li_excursion_2012}
Li B., Lam T.~Y., 2012, Monthly Notices of the Royal Astronomical Society, 425,
  730

\bibitem[{Lilje \& Lahav(1991)}]{lilje_evolution_1991}
Lilje P.~B., Lahav O., 1991, The Astrophysical Journal, 374, 29

\bibitem[{{Lilow} \& {Nusser}(2021)}]{lilow_constrained_2021}
{Lilow} R., {Nusser} A., 2021, Monthly Notices of the Royal Astronomical
  Society

\bibitem[{Lima \& Liddle(2013)}]{lima_linear_2013}
Lima N.~A., Liddle A.~R., 2013, Physical Review D, 88, 043521

\bibitem[{Linder(2005)}]{linder_cosmic_2005}
Linder E.~V., 2005, Physical Review D, 72, 43529

\bibitem[{Linder \& Cahn(2007)}]{linder_parameterized_2007}
Linder E.~V., Cahn R.~N., 2007, Astroparticle Physics, 28, 481

\bibitem[{Lombriser(2016)}]{lombriser_parametrisation_2016}
Lombriser L., 2016, Journal of Cosmology and Astro-Particle Physics, 11, 039

\bibitem[{Ma \& Bertschinger(1995)}]{ma_cosmological_1995}
Ma C.-P., Bertschinger E., 1995, The Astrophysical Journal, 455, 7

\bibitem[{Macaulay {et~al}\mbox{.}(2013)Macaulay, Wehus, \&
  Eriksen}]{Macaulay_2013}
Macaulay E., Wehus I.~K., Eriksen H.~K., 2013, Physical Review Letters, 111

\bibitem[{Magnano \& Soko\l{}owski(1994)}]{magnano_physical_1994}
Magnano G., Soko\l{}owski L.~M., 1994, Physical Review D, 50, 5039

\bibitem[{Mancinelli \& Yahil(1995)}]{mancinelli_local_1995}
Mancinelli P.~J., Yahil A., 1995, The Astrophysical Journal, 452, 75

\bibitem[{Mancinelli {et~al}\mbox{.}(1993)Mancinelli, Yahil, Canon, \&
  Dekel}]{mancinelli_nonlinear_1993}
Mancinelli P.~J., Yahil A., Canon G., Dekel A., 1993, in On nonlinear
  approximations to cosmic problems with mixed boundary conditions, p. 215

\bibitem[{Mandal \& Nadkarni-Ghosh(2020)}]{mandal_one-point_2020}
Mandal A., Nadkarni-Ghosh S., 2020, Monthly Notices of the Royal Astronomical
  Society, 498, 355

\bibitem[{Martino {et~al}\mbox{.}(2009)Martino, Stabenau, \&
  Sheth}]{martino_2009}
Martino M.~C., Stabenau H.~F., Sheth R.~K., 2009, Phys. Rev. D, 79, 084013

\bibitem[{{Monaco}(2016)}]{2016monaco}
{Monaco} P., 2016, Galaxies, 4, 53

\bibitem[{Moretti {et~al}\mbox{.}(2020)Moretti, Mozzon, Monaco, Munari, \&
  Baldi}]{Moretti_2020}
Moretti C., Mozzon S., Monaco P., Munari E., Baldi M., 2020, Monthly Notices of
  the Royal Astronomical Society, 493, 1153

\bibitem[{Nadkarni-Ghosh(2013)}]{nadkarni-ghosh_non-linear_2013}
Nadkarni-Ghosh S., 2013, Monthly Notices of the Royal Astronomical Society,
  428, 1166

\bibitem[{Nadkarni-Ghosh \& Chernoff(2011)}]{nadkarni-ghosh_extending_2011}
Nadkarni-Ghosh S., Chernoff D.~F., 2011, Monthly Notices of the Royal
  Astronomical Society, 410, 1454

\bibitem[{Nadkarni-Ghosh \& Chernoff(2013)}]{nadkarni-ghosh_modelling_2013}
Nadkarni-Ghosh S., Chernoff D.~F., 2013, Monthly Notices of the Royal
  Astronomical Society, 431, 799

\bibitem[{Nadkarni-Ghosh \&
  Refregier(2017)}]{nadkarni-ghosh_einstein-boltzmann_2017}
Nadkarni-Ghosh S., Refregier A., 2017, Monthly Notices of the Royal
  Astronomical Society, 471, 2391

\bibitem[{Nadkarni-Ghosh \& Singhal(2016)}]{nadkarni-ghosh_phase_2016}
Nadkarni-Ghosh S., Singhal A., 2016, Monthly Notices of the Royal Astronomical
  Society, 457, 2773

\bibitem[{Nojiri {et~al}\mbox{.}(2017)Nojiri, Odintsov, \&
  Oikonomou}]{Nojiri_2017}
Nojiri S., Odintsov S., Oikonomou V., 2017, Physics Reports, 692, 1–104

\bibitem[{Nojiri \& Odintsov(2007)}]{Nojiri_2007}
Nojiri S., Odintsov S.~D., 2007, Physics Letters B, 657, 238–245

\bibitem[{Nojiri \& Odintsov(2011)}]{Nojiri_2011}
Nojiri S., Odintsov S.~D., 2011, Physics Reports, 505, 59–144

\bibitem[{Nusser(2017)}]{nusser_velocity-density_2017}
Nusser A., 2017, Monthly Notices of the Royal Astronomical Society, 470, 445

\bibitem[{Nusser {et~al}\mbox{.}(2012)Nusser, Branchini, \&
  Davis}]{nusser_new_2012}
Nusser A., Branchini E., Davis M., 2012, The Astrophysical Journal, 744, 193

\bibitem[{{Nusser} {et~al}\mbox{.}(2014){Nusser}, {Davis}, \&
  {Branchini}}]{nusser_recovery_2014}
{Nusser} A., {Davis} M., {Branchini} E., 2014, The Astrophysical Journal, 788,
  157

\bibitem[{Nusser {et~al}\mbox{.}(1991)Nusser, Dekel, Bertschinger, \&
  Blumenthal}]{nusser_cosmological_1991}
Nusser A., Dekel A., Bertschinger E., Blumenthal G.~R., 1991, The Astrophysical
  Journal, 379, 6

\bibitem[{Oyaizu(2008)}]{oyaizu_nonlinear_2008}
Oyaizu H., 2008, Physical Review D, 78, 123523

\bibitem[{{Pace} {et~al}\mbox{.}(2021){Pace}, {Battye}, {Bellini}, {Lombriser},
  {Vernizzi}, \& {Bolliet}}]{pace_comparison_2021}
{Pace} F., {Battye} R.~A., {Bellini} E., {Lombriser} L., {Vernizzi} F.,
  {Bolliet} B., 2021, Journal of Cosmology and Astroparticle Physics, 2021, 017

\bibitem[{Paranjape {et~al}\mbox{.}(2013)Paranjape, Sheth, \&
  Desjacques}]{paranjape_excursion_2013}
Paranjape A., Sheth R.~K., Desjacques V., 2013, Monthly Notices of the Royal
  Astronomical Society, 431, 1503–1512

\bibitem[{Peebles(1980)}]{Peebles80}
Peebles P., 1980, The Large-Scale Structure of the Universe. Princeton
  University Press

\bibitem[{{Planck Collaboration} {et~al}\mbox{.}(2020){Planck Collaboration},
  Aghanim, Akrami, Ashdown, Aumont, Baccigalupi, Ballardini, Banday, Barreiro,
  Bartolo, Basak, Battye, Benabed, Bernard, Bersanelli, Bielewicz, Bock, Bond,
  Borrill, Bouchet, Boulanger, Bucher, Burigana, Butler, Calabrese, Cardoso,
  Carron, Challinor, Chiang, Chluba, Colombo, Combet, Contreras, Crill,
  Cuttaia, de~Bernardis, de~Zotti, Delabrouille, Delouis, Di~Valentino, Diego,
  DorÃ©, Douspis, Ducout, Dupac, Dusini, Efstathiou, Elsner, EnÃlin,
  Eriksen, Fantaye, Farhang, Fergusson, Fernandez-Cobos, Finelli, Forastieri,
  Frailis, Fraisse, Franceschi, Frolov, Galeotta, Galli, Ganga,
  GÃ©nova-Santos, Gerbino, Ghosh, GonzÃ¡lez-Nuevo, GÃ³rski, Gratton,
  Gruppuso, Gudmundsson, Hamann, Handley, Hansen, Herranz, Hildebrandt, Hivon,
  Huang, Jaffe, Jones, Karakci, KeihÃ€nen, Keskitalo, Kiiveri, Kim, Kisner,
  Knox, Krachmalnicoff, Kunz, Kurki-Suonio, Lagache, Lamarre, Lasenby,
  Lattanzi, Lawrence, Le~Jeune, Lemos, Lesgourgues, Levrier, Lewis, Liguori,
  Lilje, Lilley, Lindholm, LÃ³pez-Caniego, Lubin, Ma, MacÃ­as-PÃ©rez,
  Maggio, Maino, Mandolesi, Mangilli, Marcos-Caballero, Maris, Martin,
  Martinelli, MartÃ­nez-GonzÃ¡lez, Matarrese, Mauri, McEwen, Meinhold,
  Melchiorri, Mennella, Migliaccio, Millea, Mitra, Miville-DeschÃªnes,
  Molinari, Montier, Morgante, Moss, Natoli, NÃžrgaard-Nielsen, Pagano,
  Paoletti, Partridge, Patanchon, Peiris, Perrotta, Pettorino, Piacentini,
  Polastri, Polenta, Puget, Rachen, Reinecke, Remazeilles, Renzi, Rocha,
  Rosset, Roudier, RubiÃ±o-MartÃ­n, Ruiz-Granados, Salvati, Sandri,
  Savelainen, Scott, Shellard, Sirignano, Sirri, Spencer, Sunyaev, Suur-Uski,
  Tauber, Tavagnacco, Tenti, Toffolatti, Tomasi, Trombetti, Valenziano,
  Valiviita, Van~Tent, Vibert, Vielva, Villa, Vittorio, Wandelt, Wehus, White,
  White, Zacchei, \& Zonca}]{planck_collaboration_planck_2020}
{Planck Collaboration} {et~al.}, 2020, Astronomy and Astrophysics, 641, A6

\bibitem[{Pogosian \& Silvestri(2008)}]{pogosian_pattern_2008}
Pogosian L., Silvestri A., 2008, Physical Review D, 77, 023503

\bibitem[{Press \& Schechter(1974)}]{press_formation_1974}
Press W.~H., Schechter P., 1974, Astrophysical Journal, 187, 425

\bibitem[{Rampf \& Frisch(2017)}]{Rampf_2017}
Rampf C., Frisch U., 2017, Monthly Notices of the Royal Astronomical Society,
  471, 671

\bibitem[{Rampf \& Hahn(2020)}]{Rampf_2020}
Rampf C., Hahn O., 2020, Monthly Notices of the Royal Astronomical Society:
  Letters, 501, L71

\bibitem[{Rizzo {et~al}\mbox{.}(2017)Rizzo, Villaescusa-Navarro, Monaco,
  Munari, Borgani, Castorina, \& Sefusatti}]{Rizzo_2017}
Rizzo L.~A., Villaescusa-Navarro F., Monaco P., Munari E., Borgani S.,
  Castorina E., Sefusatti E., 2017, Journal of Cosmology and Astroparticle
  Physics, 2017, 008

\bibitem[{Ruan {et~al}\mbox{.}(2020)Ruan, Zhang, \& Hu}]{Ruan_2020}
Ruan C.-Z., Zhang T.-J., Hu B., 2020, Monthly Notices of the Royal Astronomical
  Society, 492, 4235

\bibitem[{Sawicki \& Bellini(2015)}]{sawicki_limits_2015}
Sawicki I., Bellini E., 2015, Physical Review D, 92, 084061

\bibitem[{Sch\"{a}fer \& Koyama(2008)}]{schafer_spherical_2008}
Sch\"{a}fer B.~M., Koyama K., 2008, Monthly Notices of the Royal Astronomical
  Society, 385, 411

\bibitem[{Schmidt {et~al}\mbox{.}(2010)Schmidt, Hu, \&
  Lima}]{schmidt_spherical_2009}
Schmidt F., Hu W., Lima M., 2010, Physical Review D, 81

\bibitem[{Schmidt {et~al}\mbox{.}(2009)Schmidt, Lima, Oyaizu, \&
  Hu}]{schmidt_nonlinear_2009}
Schmidt F., Lima M., Oyaizu H., Hu W., 2009, Physical Review D, 79, 83518

\bibitem[{Scoccimarro(2004)}]{scoccimarro_redshift-space_2004}
Scoccimarro R., 2004, Physical Review D, 70, 83007

\bibitem[{Sheth {et~al}\mbox{.}(2001)Sheth, Mo, \& Tormen}]{Sheth_2001}
Sheth R.~K., Mo H.~J., Tormen G., 2001, Monthly Notices of the Royal
  Astronomical Society, 323, 1–12

\bibitem[{Silvestri {et~al}\mbox{.}(2013)Silvestri, Pogosian, \&
  Buniy}]{silvestri_practical_2013}
Silvestri A., Pogosian L., Buniy R.~V., 2013, Physical Review D, 87, 104015

\bibitem[{Slotine \& Weiping(1991)}]{slotine}
Slotine J.~E., Weiping L., 1991, Applied non-linear control. Prentice-Hall
  Inc., Englewood Cliffs, New Jersey 07632 USA

\bibitem[{Song {et~al}\mbox{.}(2021)Song, Moretti, Monaco, \& Hu}]{song_2021}
Song Y., Moretti C., Monaco P., Hu B., 2021

\bibitem[{{Song} \& {Dor{\'e}}(2009)}]{yong_step_2009}
{Song} Y.-S., {Dor{\'e}} O., 2009, Journal of Cosmology and Astroparticle
  Physics, 2009, 025

\bibitem[{Song {et~al}\mbox{.}(2007)Song, Hu, \& Sawicki}]{song_large_2007}
Song Y.-S., Hu W., Sawicki I., 2007, Physical Review D, 75, 044004

\bibitem[{Sotiriou \& Faraoni(2010)}]{sotiriou_fr_2010}
Sotiriou T.~P., Faraoni V., 2010, Reviews of Modern Physics, 82, 451

\bibitem[{Starobinsky(1980)}]{starobinsky_new_1980}
Starobinsky A., 1980, Physics Letters B, 91, 99

\bibitem[{Starobinsky(2007)}]{starobinsky_disappearing_2007}
Starobinsky A.~A., 2007, Soviet Journal of Experimental and Theoretical Physics
  Letters, 86, 157

\bibitem[{Strogatz(1994)}]{strogatz}
Strogatz S.~H., 1994, Non-linear dynamics and chaos. Perseus Publishing,
  Cambridge, Massachusets

\bibitem[{Susperregi \& Buchert(1997)}]{susperregi_cosmic_1997}
Susperregi M., Buchert T., 1997, Astronomy and Astrophysics, 323, 295

\bibitem[{Uzan(2007)}]{uzan_acceleration_2007}
Uzan J.-P., 2007, General Relativity and Gravitation, 39, 307, arXiv:
  astro-ph/0605313

\bibitem[{Valogiannis {et~al}\mbox{.}(2020)Valogiannis, Bean, \&
  Aviles}]{Valogiannis_2020}
Valogiannis G., Bean R., Aviles A., 2020, Journal of Cosmology and
  Astroparticle Physics, 2020, 055?055

\bibitem[{Wolfram~Research(2018)}]{mathematica}
Wolfram~Research I., 2018, Mathematica, version 11.3.0 edn. Wolfram Research,
  Inc., Champaign, Illinois

\bibitem[{Zaroubi {et~al}\mbox{.}(1999)Zaroubi, Hoffman, \&
  Dekel}]{zaroubi_wiener_1999}
Zaroubi S., Hoffman Y., Dekel A., 1999, The Astrophysical Journal, 520, 413

\bibitem[{Zel'Dovich(1970)}]{zeldovich_gravitational_1970}
Zel'Dovich Y.~B., 1970, Astronomy and Astrophysics, 5, 84

\end{thebibliography}

\appendix
\section{Initial Profiles}
\label{app:tophatsetup}
The compensated top-hat is a one-dimensional function represented as
\bea 
\delta_{top}(x) &=& A\;\;\; \;\;\; \;\;\;\; 0< x \leq x_{top}\\
&=& -1 \;\;\;  x_{top} < x \leq x_u \\
&=& 0 \;\;\; {\rm otherwise,}
\eea
where $A$ is the amplitude of the top-hat. $x_{top}$ and $x_u$ are the boundaries of the overdense region and the underdense compensating region and are related to through conservation of mass as 
\beq
(1+A) x_{top}^3 = x_u^3. 
\eeq
A smooth top-hat is obtained by the following function 
\beq 
\delta_{smooth}(x) = \frac{1}{\sqrt{2 \pi \sigma^2 x^2}} \left\{\int_0^{x_{top}} A \left(e^{-\frac{(x-y)}{2 \sigma^2}} - e^{-\frac{(x+y)}{2 \sigma^2}}\right) y dy + \int_{x_{top}}^{x_u} (-1) \left(e^{-\frac{(x-y)}{2 \sigma^2}} - e^{-\frac{(x+y)}{2 \sigma^2}}\right) y dy \right\}, 
\eeq
where $\sigma$ is the smoothing parameter. In this paper we have used two different profiles with $\sigma = 0.1$ and $\sigma = 0.0025$. In the numerical implementation, $L_{min}$ and $L_{max}$ denote the minimum and maximum values of the grid points along the $x$-direction. The value of $A$ and $x_{top}$ are chosen according to the calculation performed. $N_{t}$ gives the number of time steps used to divide the time interval from the initial epoch $a_{switch}$ to the final epoch $a_{final}$. $N_s$ denotes the number of spatial grid points, equi-spaced in $\ln x$ along the radial direction. Tables \ref{Table1} and \ref{Table2} give the list of parameters used in different sections of the paper. 

\begin{table}
 \begin{tabular}{|c|c|c|c|c|c|}
    \hline
     \textbf{Profile Name} & $\sigma$ & \textbf{$x_{top}$} & $Q$ (for $f_{R0} =-10^{-6}$) & \textbf{$L_{min}$} ($h^{-1}$Mpc) & \textbf{$L_{max}$} ($h^{-1}$Mpc)\\
      \hline
      1a &  0.0025 & 0.002    &1217 	&	0.0001	&	0.1  \\
     1b &  0.0025 & 2	         & 1.21     &	0.1		&	100  \\
     1c &  0.0025 & 200  	 & 0.012 	&	10		&	10000  \\
     \hline
       1d & 0.0025 &0. 02     & 121 	& 0.001 		&      1\\
      1e & 0.0025 &0.1     & 24.34 	& 0.005 		&      5\\
      1f & 0.0025 &0.2     & 12.1  	& 0.01 		&      10\\
      1g & 0.0025 &1       & 2.43	 	& 0.05 		&      50\\
      1h & 0.0025 &10     & 0.24	 	& 0.5 		&      500\\
      1i & 0.0025 &20      & 0.12		& 1		        &      1000\\
      1j & 0.0025 &100    & 0.024	& 5		        &      5000\\
     \hline
     2a &  0.1	 & 0.002      & 1217	&	0.0001	&	0.1  \\
     2b &  0.1	 & 2	  	 &1.21    &	0.1		&	100  \\
     2c &  0.1 	& 200	& 0.012        &	10		&	10000  \\
     2d &  0.1 	& 0.02	& 121        &	0.001		&	1  \\
     \hline
    \end{tabular}
       \caption{Table denoting the parameters of the two profiles used. The profiles 1e-1j are used only in \figref{psiratio} .}
       \label{Table1}
\end{table}

\begin{table}
    \begin{tabular}{|c|c|c|c|c|c|c|}
    \hline
     \textbf{Figure} & Profile used &$A$ & $N_s$ & $N_t$ & $a_{switch}$ & $a_{final}$\\
      \hline
    \figref{staticsoln}  &   1a,b,c                  	 &  7	         	   		 		&   3000  		     &    -		  &	-	&	-  \\
     \figref{psiratio}  &    1b-1j                       &  7	         	   		 		&   3000  		     &    -		  &	-	&	-  \\
    \figref{lindelta}     &   1a and 1b 	               	 &  0.0007 	    				&   500  		    &   40   		  & 0.1	&	0.2, 0.4   \\
    \figrefs{denvel}, \figrefbare{phasespace}, \figrefbare{deviation}      &   1a,b,c and 2a,b,c & 0.0008	            				&  1600        	    &  320  		  &  0.1 	&	1  \\    
    \figref{fit} 	&   1a                		&  0.0008			                           & 1600  		    & 320             &  0.1  	&       1\\
    \figref{highNL} 	&   1a and 1c 		&  0.0015 (for 1c) 0.002 (for 1d)       & 1600  		    & 320             &  0.1  	&       1\\
    \figref{figerror}     & 1d                         & 0.0008					        &  $5 \times N_t$  &  20  to 640  &  0.1 	&	1  \\
    \hline
        \end{tabular}
       \caption{Table characterizing the initial data used in various figures in the text. \capfigrefs{staticsoln} and \figrefbare{psiratio} correspond to static solutions and there are no parameters for the temporal evolution}.    
       \label{Table2}
\end{table} 
We found that it was numerically more stable to use compensated top-hats because the potentials $\Phi$ and $\Psi$ analytically vanish after a finite extent. For a pure top-hat or a more realistic profile such as the one based on peaks theory \citep{bardeen_statistics_1986,lilje_evolution_1991} and used in \cite{nadkarni-ghosh_non-linear_2013} or \cite{kopp_spherical_2013} is not as stable because the fields are non-zero at any finite radius (which is inevitable in the discretization). 

\section{Convergence tests}
\label{app:error}
\begin{figure}
\includegraphics[width=8cm]{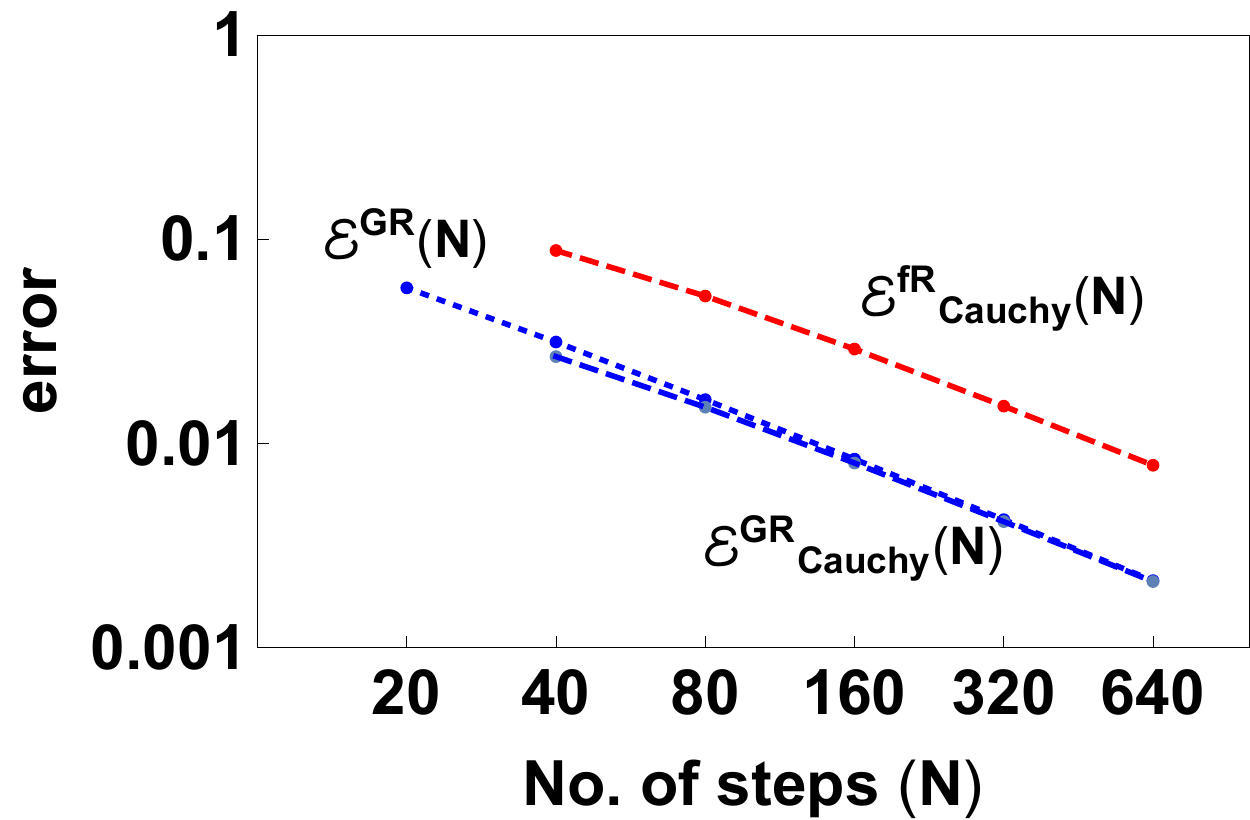}  
\caption{Convergence test of the algorithm. Comparison of successive approximations for evolution from $a = 0.1$ to $a=1$ for profile 1a for the variable $\Delta$. The algorithm was applied to a $\Lambda$CDM model with $\Phi = \Psi$ and an $f(R)$ model with $f_{R0} = -10^{-6}$.  }
\label{figerror}
\end{figure}

The algorithm outlined in \S \ref{sec:algorithm} was used to evolve an initial profile from $a=0.001$ to $a=1$ with the transition from $f(R)$ to GR at $a_{switch}=0.1$. The evolution from $a=0.1$ to $a=1$ was carried out using successive runs with $N_t = 20, 40, 80, 160, 320$ and 640 steps. For each run, the spatial domain was divided into $N_s$ steps equi-spaced in the $\ln x$ direction. In order to achieve convergence, the spatial domain needs to be refined along with the refinement in the temporal domain. We choose $N_s = 5 \times N_t$. All solutions to the second order ordinary differential equations (O.D.E.s) we computed using the inbuilt function {\it NDSolve} in the software package MATHEMATICA \citep{mathematica}.  

In the GR case, the exact answer can be computed in a single step. As a code test, we computed approximate answers for the GR case using the same algorithm but assuming $\Phi = \Psi$ and compared to the exact answer as well as compared successive approximations to give the Cauchy error. For the $f(R)$ evolution we compared only successive approximations. 
We define the errors as follows 
\bea
{\mathcal E}(N) &=& \sqrt{\frac{1}{N} \sum_{i=1}^{N_s} |f_N(x_i) - f_{exact}(x_i)|^2}\\
{\mathcal E}_{Cauchy}(N) &=& \sqrt{\frac{1}{N} \sum_{i=1}^{N_s} |f_N(x_i) - f_{2N}(x_i)|^2}, 
\eea
where $f$ denotes the function to be compared. The first error cannot be computed in the $f(R)$ case for lack of knowledge of the exact solution. The spatial domain has a different grid spacing in each approximation. Thus to compare, we need to interpolate the approximation which has the finer grid and compute it at the domain points corresponding to the coarser grid. 
\capfigref{figerror} plots the errors for both GR and $f(R)$ in the evolved average density $\Delta$ for the profile 1d with the parameters as shown in table \ref{Table2}. We found the trends to be the same for the variables $\delta$ and $v$ for profile 1e and the same trends for all three variables for profile 2d. 

\section{Derivation of the analytical solution}
The equation for $\Phi_+$ is the usual Poisson equation with the solution given by 
\beq
\Phi_+(x) = C_1 + \int_{k_1}^x \left\{\frac{C_2}{z^2}+ \frac{1}{z^2} \int_{k_2}^z  {\mathcal P}_1(a) \delta(y) y^2 dy \right\} dz, 
\eeq  
where $y$ and $z$ are dummy variables for the integrals and 
\beq 
{\mathcal P}_1(a) = \frac{3 \Omega_m(a) a^2}{2}.
\eeq
$k_1$ and $k_2$ are arbitrary constants which are related to the boundary points along the $x$ axis. Imposing the boundary condition at $x= \infty$, gives 
\beq 
C_1= - \int_{k_1}^\infty \left\{\frac{C_2}{z^2}+ \frac{1}{z^2} \int_{k_2}^z {\mathcal P}_1(a) \delta(y) y^2 dy \right\} dz, 
\eeq
 The derivative 
\beq 
\left.\nabla_x {\tilde \Phi_+}\right|_{x=0}  = \frac{C_2}{x^2} + \frac{1}{x^2} \int_{k_2}^y {\mathcal P}_1(a) \delta(y) y^2 dy =0 \implies C_2 =  \int_0^{k_2}{ \mathcal P}(a) \delta(y). y^2 dy 
\eeq
 Thus, ${\tilde \Phi_+}$ and its derivative (force) are  
 \bea
 {\tilde \Phi_+} (x) &=& {\mathcal P}_1(a)  \int_\infty^x   \frac{1}{z^2} \int_0^z \delta(y) y^2 dy =  \frac{{\mathcal P}_1(a)}{3} \int_\infty^x  z \Delta(z) dz\\
 \nabla_x {\tilde \Phi_+} &=& {\mathcal P}_1(a)  \frac{1}{x^2} \int_0^x \delta(y) y^2 dy = \frac{{\mathcal P}_1(a)}{3} x \Delta(x), 
 \eea
 where the spherically averaged fractional density contrast $\Delta$ is defined as: 
 \beq 
 \Delta(x) = \frac{3}{x^3} \int_0^x \delta(y) y^2 dy. 
 \eeq
 The general solution of \eqnref{chieq} for ${\tilde \chi}$
\beq 
{\tilde \chi}(x) = C_1 \frac{e^{-\sqrt{{\mathcal P}_2} x}}{x} + \frac{C_2}{2 \sqrt{{\mathcal P}_2}}  \frac{e^{\sqrt{{\mathcal P}_2}x}}{x}+ \frac{{{\mathcal P}_3}}{2\sqrt{{\mathcal P}_2}} \frac{e^{-\sqrt{{\mathcal P}_2}x}}{ x} \int_{k_1}^x e^{\sqrt{{\mathcal P}_2}y} y\delta(y) dy - \frac{{{\mathcal P}_3}}{2\sqrt{{\mathcal P}_2}} \frac{e^{\sqrt{{\mathcal P}_2}x}}{x} \int_{k_2}^x e^{-\sqrt{{\mathcal P}_2}y} y\delta(y) dy, 
\eeq 
where 
\beq
{{\mathcal P}_2} = \frac{1}{{\bar x}_C^2} \; \;\; {\rm and} \;\;\; {{\mathcal P}_3} = a^2 \Omega_m(a). 
\eeq
At $x = \infty$, the first and third terms vanish since $\delta(x)$ has a finite extent. Thus, imposing the boundary condition at $x= \infty$ gives 
\beq 
C_2 = {{\mathcal P}_3}\int_{k_2}^\infty e^{-\sqrt{{\mathcal P}_2}y} y\delta(y) dy.
\eeq
This gives 
\beq
{\tilde \chi}(x) = C_1 \frac{e^{-\sqrt{{\mathcal P}_2}x}}{x} + 
\frac{{\mathcal P}_3}{2 \sqrt{{\mathcal P}_2}x} \left[ e^{\sqrt{{\mathcal P}_2}x} L_2(k_2,\infty) +e^{-\sqrt{{\mathcal P}_2}x} L_1(k_1,x) - e^{\sqrt{{\mathcal P}_2}x}  L_2(k_2,x)
\right]
\eeq
where 
\bea
L_1(k_1,x)  &=& \int_{k1}^x  e^{\sqrt{{\mathcal P}_2}y} y\delta(y) dy  \\
L_2(k_2,x) &=& \int_{k_2}^x   e^{-\sqrt{{\mathcal P}_2}y} y\delta(y) dy
\eea
Taking the derivative 
\bea
\frac{d {\tilde \chi}}{dx} = &-&\frac{C_1 e^{-\sqrt{{\mathcal P}_2}x}}{x^2} (1 + \sqrt{{\mathcal P}_2} x) \\
&+& \frac{{\mathcal P}_3}{2 \sqrt{{\mathcal P}_2}x^2 }\left[ L_2(k_2, \infty) (\sqrt{\mathcal P}_2 x -1) e^{\sqrt{\mathcal P}_2 x} - L_1(k,x) (1 + \sqrt{\mathcal P}_2 x ) e^{-\sqrt{\mathcal P}_2 x} - L_2(k,x) (\sqrt{\mathcal P}_2 x -1 ) e^{\sqrt{\mathcal P}_2 x}
\right]
\eea
Setting  $\left.\frac{d {\tilde \chi}}{dx}\right|_{x \approx 0} = 0$ gives 
\beq 
C_1 = - \frac{{\mathcal P}_3}{2 \sqrt{{\mathcal P}_2}} \left[L_1(k_1,0) - L_2(k_2,0) + L_2(k_2, \infty) \right]
\eeq
Substituting for $C_1$ and $C_2$, writing  
\beq  -L_2(k_2,0) + L_2(k_2, \infty) = \int_0^\infty e^{-\sqrt{{\mathcal P}_2}y} y\delta(y) dy =  \int_0^x e^{-\sqrt{{\mathcal P}_2}y} y\delta(y) dy +  \int_x^\infty e^{-\sqrt{{\mathcal P}_2}y} y\delta(y) dy
\eeq
and rearranging the terms gives 
\bea
 {\tilde\chi}(x) &=& \frac{\mathcal P_3 }{\sqrt{\mathcal P_2}}\frac{1}{x} \left[ \exp(-\sqrt{\mathcal P_2}x) I_1(x) + \sinh(\sqrt{\mathcal P_2}x) I_2(x) \right],\\
 \label{gradchi}\nabla_x {\tilde \chi} &=& - \frac{{\tilde \chi} }{x}+ \frac{{\mathcal P_3}}{x} \left[-\exp(-\sqrt{\mathcal P_2}x) I_1(x) + \cosh(\sqrt{\mathcal P_2}x)  I_2(x) \right]. 
 \eea
 where 
 \bea
 I_1(x) &=&  \int_0^x   \left\{ \sinh (\sqrt{\mathcal P_2}y) \delta(y) y \right\} dy,\\
I_2(x) &=& \int_x^\infty \left\{\exp(-\sqrt{\mathcal P_2}y)\delta(y)   y \right\} dy.
 \eea

\end{document}